\documentclass[lettersize,journal]{IEEEtran}

\usepackage{array}
\usepackage[caption=false,font=normalsize,labelfont=sf,textfont=sf]{subfig}
\usepackage{url}
\usepackage{verbatim}
\usepackage{graphicx}
\usepackage[hidelinks]{hyperref}
\usepackage{amsmath,amssymb,amsfonts}
\usepackage{algpseudocode}
\usepackage[linesnumbered,ruled,vlined]{algorithm2e}
\usepackage{pifont}
\usepackage{booktabs}
\usepackage{graphicx}
\usepackage{textcomp}
\usepackage{xcolor}
\usepackage{pgf}
\usepackage{tikz}
\usepackage{color}
\usepackage{stfloats}
\usepackage{multirow}
\usepackage{diagbox}
\usepackage{url}
\usepackage{subcaption}
\usepackage{parallel}
\usepackage{float}
\usepackage{placeins}

\usepackage{fancyhdr}
\usepackage[normalem]{ulem}
\usepackage{cite}

\newcommand{\squishlist}{
   \begin{list}{$\bullet$}
    { \setlength{\itemsep}{0pt}      \setlength{\parsep}{0pt}
      \setlength{\topsep}{3pt}       \setlength{\partopsep}{0pt}
      \setlength{\listparindent}{-2pt}
      \setlength{\itemindent}{-5pt}
      \setlength{\leftmargin}{1em} \setlength{\labelwidth}{0em}
      \setlength{\labelsep}{0.5em} } }

\newcommand{\squishend}{
    \end{list}  }

\newcommand*\blackcircledempty[1]{\tikz[baseline=(char.base)]{
        \node[shape=circle, text={rgb,255:red,0;green,0;blue,0}, font=\small, draw={rgb,255:red,0;green,0;blue,0},inner sep=0.5pt] (char) {#1};}}

\newif\ifshowreviewtags
\showreviewtagstrue

\begin{document}

\title{ESR-HGNN: Eliminating Semantic Redundancy\\ for Efficient Mini-batch HGNN Inference}

\author{
        Dengke~Han,
        Mingyu~Yan,~\IEEEmembership{Senior Member,~IEEE},
        Duo~Wang,
        Wenming~Li, \\
        Xiaochun~Ye,
        and~Dongrui~Fan,
        ~\IEEEmembership{Senior~Member,~IEEE}
\IEEEcompsocitemizethanks{
\IEEEcompsocthanksitem

This work was supported in part by the Beijing Nova Program under Grant No. 20250484774, the CAS Project for Young Scientists in Basic Research under Grant No. YSBR-029, and the CAS Project for Youth Innovation Promotion Association.

Dengke Han, Mingyu Yan, Duo Wang, Wenming Li, Xiaochun Ye, and Dongrui Fan are with the State Key Lab of Processors, Institute of Computing Technology, Chinese Academy of Sciences, Beijing 100045, China, and the University of Chinese Academy of Sciences, Beijing 101408, China (e-mail: {handengke21s, yanmingyu, wangduo, liwenming, yexiaochun, fandr}@ict.ac.cn). Mingyu Yan is the corresponding author.

}
}

\markboth{IEEE Transactions on Parallel and Distributed Systems}
{Shell \MakeLowercase{\textit{et al.}}: Bare Demo of IEEEtran.cls for Computer Society Journals}


\maketitle

\begin{abstract}

Heterogeneous graph neural networks (HGNNs) are highly effective in processing heterogeneous graph data and have been widely adopted in critical domains. As real-world graph data continues to scale, performing direct inference on entire graphs becomes increasingly infeasible, making mini-batch methods the standard approach. However, in end-to-end HGNN inference, metapath-based mini-batch sampling constitutes a significant performance bottleneck due to the extensive random memory accesses induced by the irregular traversal of graph structures. Existing sampling paradigms suffer from excessive redundant traversals caused by inherent semantic redundancy, severely degrading sampling efficiency and, consequently, leading to suboptimal mini-batch inference performance.

In this work, we propose a redundancy-aware HGNN sampling paradigm that leverages a metapath trie to reuse traversal paths, effectively eliminating redundant memory accesses. We then map it onto a multi-channel hardware sampling unit denominated ESR-HGNN. Furthermore, we introduce a reusability-driven metapath grouping technique that optimally clusters metapaths to maximize reusable traversal paths within hardware channels, enhancing efficiency in scenarios with semantic parallelism. Extensive experimental results demonstrate that ESR-HGNN achieves an average sampling performance improvement of one order of magnitude over CPU and GPU, accompanied by significant energy savings. Additionally, it delivers substantial speedup in end-to-end mini-batch inference when integrated with GPU and state-of-the-art HGNN inference accelerator.


\begin{IEEEkeywords}
Heterogeneous Graph Neural Network, Mini-batch Sampling, HGNN Accelerator, Redundancy Elimination
\end{IEEEkeywords}

\end{abstract}

\section{Introduction}

Graph Neural Networks (GNNs) have demonstrated remarkable power in processing non-Euclidean data in recent years, leading to their widespread application across critical domains. Early advancements in GNNs centered on homogeneous graphs (HomoGs), consisting of a single type of vertex and edge. However, much real-world data naturally takes the form of heterogeneous graphs (HetGs), which contain multiple types of vertices and edges. Unlike GNNs designed for HomoGs, Heterogeneous Graph Neural Networks (HGNNs) can extract not only structural information but also semantic insights embedded in diverse relations. With their robust representational capabilities, HGNNs have become indispensable in various fields, including recommendation systems~\cite{weibo-recommendation, dataset-recommendation}, cybersecurity~\cite{tencent_malware_detection, abnormal_event_detection}, electronic design automation~\cite{eda_1, eda_2}, and many others.


As real-world graph data continues to scale rapidly, performing HGNN inference on entire graphs has become increasingly impractical due to hardware limitations, necessitating the adoption of the mini-batch execution paradigm~\cite{GraphSage,sampling_survey,Comprehensive_Survey_GNN_Distributed_Training}. This paradigm involves sampling the neighbors of target vertices from the original graph, which determines the workload for subsequent inference. As many leading HGNN models~\cite{HAN, HPN, MAGNN} utilize metapaths to capture diverse semantic information, HGNN sampling requires multi-semantic neighbor traversal, where each traversal path follows a specific sequence of distinct relations. This process exhibits highly irregular memory access patterns, resulting in significant performance bottlenecks across a variety of scenarios including both single-node and distributed systems~\cite{understand_hgnn_training}.


The expanding scale of graph data, coupled with an increasing diversity of vertex types, has naturally led to the more frequent adoption of longer metapaths. Additionally, previous work~\cite{SeHGNN} demonstrates that longer metapaths effectively expand the receptive field, thus improving the accuracy of HGNN models. However, the incorporation of longer metapaths introduces increased complexity in the sampling process, accentuating the irregularity of memory access patterns. Specifically, longer metapaths necessitate the traversal of a greater number of distinct adjacency relations, amplifying random memory access during the sampling process. As the length of metapaths increases, the overlap between longer metapaths and their shorter counterparts becomes more pronounced, a phenomenon referred to as semantic redundancy. This redundancy leads to a higher frequency of redundant traversal paths across various metapaths during mini-batch sampling, rendering a substantial redundant memory accesses.



Traditional computational platforms, such as CPUs and GPUs, are inherently ill-suited for the highly irregular memory access patterns associated with HGNN sampling, rendering them inefficient for these tasks. The primary challenge stems from the non-sequential memory accesses required during multi-semantic neighbor traversals. CPUs, optimized for sequential memory accesses, struggle to handle these irregular memory accesses, resulting in significant latency and suboptimal performance. While GPUs excel at parallel processing, they are primarily optimized for regular and predictable memory access patterns typically seen in conventional deep learning tasks such as matrix multiplications. As a result, the irregular memory access patterns inherent in HGNN sampling prevent the efficient utilization of GPU parallelism. Existing HGNN accelerators~\cite{MetaNMP, HiHGNN, ADE-HGNN, TLV-HGNN, GDR-HGNN} primarily optimize inference while overlooking the increasingly critical metapath-based sampling process. Consequently, inference-only optimization is insufficient to substantially improve end-to-end mini-batch performance.

To address these challenges, this work introduces a novel solution that eliminates semantic redundancy to reduce random memory accesses during the sampling process, thereby improving the overall performance of end-to-end mini-batch HGNN inference. We first propose a novel redundancy-aware sampling paradigm that dynamically records and retrieves traversal paths, enabling the efficient reuse of previously accessed paths. Subsequently, we design a multi-channel sampling unit specifically tailored for efficient hardware implementation. Furthermore, we introduce a reusability-driven semantic grouping strategy to maximize the reuse of traversal paths within groups, while simultaneously exploiting semantic parallelism to enhance the parallel execution of the sampling process. We summarize our contributions as follows:\par
\squishlist
\item
We conduct a quantitative analysis of the mini-batch HGNN inference, revealing the acceleration opportunity brought by semantic redundancy.
\item
We propose a redundancy-aware HGNN sampling paradigm to reuse redundant traversal paths and design a multi-channel sampling unit named ESR-HGNN to support the sampling method efficiently.
\item 
We present a reusability-driven grouping method for metapaths to enhance the efficacy of the sampling method in scenarios with semantic parallelism.
\item
Our comprehensive experiments demonstrate that ESR-HGNN achieves an average sampling performance improvement of 39.66$\times$ over the CPU and 13.94$\times$ over the GPU, while reducing energy consumption by 98.45\% and 86.41\%, respectively. Leveraging ESR-HGNN, the GPU and the state-of-the-art (SOTA) HGNN accelerator HiHGNN~\cite{HiHGNN} achieve a 2.78$\times$ and 5.70$\times$ speedup in end-to-end mini-batch inference compared with the CPU+GPU framework.

\squishend

\section{Background and Related Work}

\subsection{Heterogeneous Graph and Metapath}
In contrast to HomoGs, HetGs encompass multiple types of vertices and edges, embodying both structural and semantic information. A HetG is defined as $G=(V,E,\mathcal{S}^v,\mathcal{S}^e)$~\cite{SeHGNN,Simple-HGN} using notations in Table~\ref{tb:notation}, where $V$ is the set of vertices with a vertex type mapping function $\phi:V\rightarrow\mathcal{S}^v$, and $E$ is the set of edges with an edge type mapping function $\psi:E\rightarrow\mathcal{S}^e$.
Each vertex $v_i{\in}V$ is attached with a vertex type $T_v{=}\phi(v_i){\in}\mathcal{S}^v$. Each edge $e_{u, v}{\in}E\,$ is attached with a relation $R_{c_u,c_v}{=}\psi(e_{u,v}){\in}\mathcal{S}^e$, starting from the source vertex $u$ to the target vertex $v$.  A graph is heterogeneous when $|\mathcal{S}^v|+|\mathcal{S}^e|>2$, otherwise it is homogeneous.

\begin{table}[!t]
\centering
\caption{Notations and corresponding explanations.}
\label{tb:notation}
\resizebox{0.48\textwidth}{!}{
\tabcolsep=4pt
\begin{tabular}{cc|cc}
\toprule
Notation                & Explanation                           & Notation                  & Explanation       \\ \midrule
$G$                     & heterogeneous graph                   & $V$                       & vertex set \\
$E$                     & edge set                              & $\mathcal{S}^v$           & vertex type set \\
$\mathcal{S}^e$         & edge type set                         & $u,\,v$                   & vertex \\
e ($e_{u,v}$)           & edge (from $u$ to $v$)                & $R_i$ & relations \\
$T_v$                   & vertex type                         & $N_v$           & neighboring set \\

\bottomrule
\vspace{-15pt}
\end{tabular}}
\end{table}

\begin{figure*}[!ht] 
	\centering
	\vspace{-5pt}
	\includegraphics[width=0.96\textwidth]{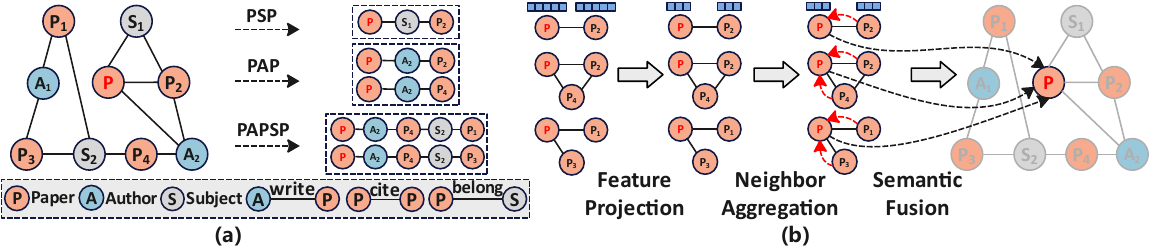}
\caption{The illustration of HetG sampling process and HGNNs.}
        \vspace{-15pt}
	\label{fig:HetG}
\end{figure*}

Fig.~\ref{fig:HetG}(a) illustrates a simple example of a HetG from the ACM dataset, which includes three types of vertices, A (Author), P (Paper), and S (Subject), along with three types of adjacency relations between them: author$\xrightarrow{\rm writes}$paper, paper$\xrightarrow{\rm cites}$paper, and paper$\xrightarrow{\rm belongs\ to}$subject (abbreviated as AP, PP and PS). Each type of relation represents a unique semantic information between the two endpoints connected. Building upon direct relations such as PS, various combinations of them can form higher-order relations, referred to as metapaths. A metapath is defined as a path in the form of $T_v^1 \xrightarrow{R_1} T_v^2 \xrightarrow{R_2} \dots \xrightarrow{R_{l-1}} T_v^l$, abbreviated as $T_v^1T_v^2\dots T_v^l$, where $T_v^i$ denotes the vertex type at the $i$-th position, $R_i$ denotes the relation between $T_v^i$ and $T_v^{i+1}$, and $l$ denotes the metapath length under our notation. A metapath represents a composite relation $R=R_1 \circ R_2 \circ \dots \circ R_{l-1}$ between the endpoint types $T_v^1$ and $T_v^l$, where $\circ$ denotes relation composition. For example, in Fig.~\ref{fig:HetG}, PSP consists of PS and SP, indicating that two papers share a subject and are thus likely related to the same research area. In general, metapaths capture richer semantic information than direct relations.

\subsection{Heterogeneous Graph Sampling}
\label{sec:background_sampling}

Graph sampling was initially introduced to enhance the convergence speed of models during GNN training and to reduce the on-the-fly hardware demands. This technique can primarily be classified into three categories: node-wise sampling, layer-wise sampling, and subgraph-wise sampling~\cite{sampling_survey}. Among these, \textit{RandomWalk}~\cite{randomwalk} is a widely used and commonly applied sampling method. It constructs a walk path by traversing neighboring vertices based on predefined traversal criteria, ultimately generating a sampled mini-batch.

Metapath2Vec~\cite{metapath2vec} is the first to incorporate the \textit{RandomWalk}~\cite{randomwalk} method into HetG sampling. This technique is a simple yet effective metapath-based sampling approach that captures complex semantic dependencies between different types of vertices and has since become the de facto standard method for metapath-based HGNN sampling. Specifically, given a metapath $T_v^1T_v^2\dots T_v^l$, for a target vertex $v$ with $\phi(v)=T_v^1$, we generate $n$ sampling paths. Let $v_j^i$ denote the vertex at the $i$-th position of the $j$-th sampling path, where $v_j^1=v$ for $j=1,\ldots,n$. The valid neighbors at the next position are defined as $N_{i+1}(v_j^i)=\{u \mid e_{v_j^i,u}\in E,\,\psi(e_{v_j^i,u})=R_i,\,\phi(u)=T_v^{i+1}\}$. At each step $i$, a valid neighbor is selected uniformly from this set. The transition probability is defined as follows:
$$
{
p(v_j^{i+1}=u \mid v_j^i) =
\begin{cases}
\frac{1}{|N_{i+1}(v_j^i)|} & u \in N_{i+1}(v_j^i) \\
0 & \text{otherwise.}
\end{cases}
}
$$
This formulation follows the relation and vertex-type sequence specified by the metapath and selects each valid neighbor with equal probability. This formula essentially indicates that each step of the HetG sampling process follows the relations of the pre-defined metapath, which is quite more complex than traditional GNN sampling~\cite{sampling_survey,GNNSampler}. To give an example, as illustrated in Fig.~\ref{fig:HetG}(a), we perform neighbor sampling for the highlighted target vertex \textit{P} in the graph using three metapaths: PSP, PAP, and PAPSP, generating the traversal paths depicted in the figure, where the endpoint of each sequence represents the neighbor sampled according to the corresponding metapath. The mini-batch is constructed based on the sampled neighbors to support the subsequent inference process.

The main differences between metapath-based HGNN sampling and traditional GNN sampling are as follows: \textbf{(a)} \textit{Multiple Semantics}: Different metapaths represent distinct complex semantics. The sampling tasks of different semantics have no data dependencies and therefore exhibit semantic parallelism. \textbf{(b)} \textit{Multiple Adjacency Relations}: Each traversal sequence under a specific semantic needs to consider various adjacency relations, significantly increasing both the complexity and workload of random memory access operations. In contrast, traditional GNN sampling involves only a single type of semantic and adjacency relation.

\subsection{Heterogeneous Graph Neural Network}

To capture both the structural and semantic information in HetGs, most prevalent HGNN models contain four primary execution stages~\cite{understand_HGNN,GNN_Characterization_Survey} as illustrated in Fig.~\ref{fig:HetG}(b). \blackcircledempty{1} \textit{Semantic Graph Build} stage builds semantic graphs for the following stages by partitioning the original HetG into a set of semantic graphs based on predefined semantics. However, in the mini-batch inference paradigm, since inference does not require processing the entire semantic graph, this execution stage is effectively substituted by the sampling process and has been omitted in the figure.
\blackcircledempty{2} \textit{Feature Projection} stage transformes the feature vector of each vertex to a new one using a multi-layer perceptron within each semantic graph.
\blackcircledempty{3} \textit{Neighbor Aggregation} stage performs the aggregation of features from neighbors within each semantic graph.
\blackcircledempty{4} \textit{Semantic Fusion} stage fuses the semantic information obtained from all semantic graphs, aiming to combine the results of the \textit{Neighbor Aggregation} stage across different semantic graphs.

\subsection{Mini-batch HGNN Inference}

\begin{figure}[!t] 
	\centering
	\includegraphics[width=0.48\textwidth]{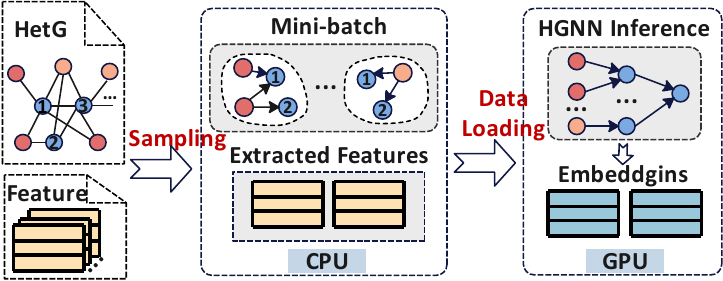}
	\vspace{-5pt}
	\caption{Workflow of HGNN mini-batch inference.}
        \vspace{-8pt}
	\label{fig:minibatch_inference}
\end{figure}

As real-world graph scales, performing inference on the entire graph simultaneously becomes impractical due to hardware constraints. As illustrated in Fig.~\ref{fig:minibatch_inference}, mini-batch inference addresses this challenge by partitioning target vertices into multiple groups. Each group undergoes sampling, followed by inference on the corresponding mini-batch. This process constitutes a single iteration, and after multiple iterations, inference for all target vertices is completed. The mini-batch inference paradigm not only mitigates hardware resource demands but also reduces inference latency for individual target vertices.

Contemporary mini-batch inference platforms primarily employ heterogeneous architectures consisting of a host CPU and a GPU as the computational nodes as in Fig.~\ref{fig:minibatch_inference}. Within this framework, the host CPU first performs sampling on the original HetGs based on predefined metapaths, generating mini-batches of target vertices. It then extracts the features corresponding to these vertices from the sampling results and transfers both the features and the mini-batch structure to the GPU via a bus, a process referred to as data loading. The GPU then executes the complete inference process of the HGNN model and returns the computed vertex embeddings. Furthermore, due to its robust parallel processing capabilities and high bandwidth, the use of GPUs for sampling has also increasingly become a widely adopted execution paradigm.

\subsection{Related Work}

\subsubsection{GNN Accelerators}
Owing to the widespread adoption of GNNs across critical domains, GNN accelerators have attracted significant interest from the architecture community in recent years~\cite{HyGCN, REFLIP-HUAKE, ViTeGNN, GraphAgile, GRIP, MultiGCN, FlowGNN, ReGNN, CoGNN, GROW, igcn}. Among these efforts, CoGNN~\cite{CoGNN} enhances mini-batch inference in traditional GNNs by utilizing reuse-aware sampling, which strategically prioritizes target vertices with a higher number of common neighbors to optimize data reuse. However, the sampling process in HGNNs involves multi-semantic neighbor traversal across various adjacency relations, resulting in a fundamentally different workflow compared with traditional GNNs. Consequently, CoGNN, along with other efforts optimized for traditional GNN sampling, is not directly applicable to the sampling process in HGNNs due to the distinct characteristics of the latter's multi-semantic neighbor traversal and complex adjacency relations.

\subsubsection{HGNN Accelerators}
Only a limited number of work~\cite{MetaNMP, HiHGNN, ADE-HGNN, GDR-HGNN} have addressed inference acceleration for emerging HGNNs, and none have specifically targeted the critical mini-batch inference scenario, where the sampling process serves as the primary performance bottleneck. Therefore, existing HGNN acceleration efforts can only achieve marginal performance improvements in the mini-batch inference scenario. In contrast, this work capitalizes on the unique multi-semantic characteristics of HGNNs to eliminate semantic redundancy, leading to significant performance enhancements in both the sampling process and end-to-end mini-batch HGNN inference.

\section{Motivation}
\label{sec:motivation}

This section first presents a quantitative analysis to identify the sampling phase as the primary performance bottleneck in end-to-end mini-batch inference. We further examine the execution behavior characteristics of the sampling phase, investigate opportunities for accelerating this phase through the elimination of semantic redundancy, and highlight the challenges encountered within traditional computational platforms. The experimental configurations are identical to those described in Section~\ref{sec:experiment_setup}. The CPU sampling baseline employs 32 worker threads, one per physical core. A larger thread count is not adopted because additional threads increase shared-cache and memory-access contention, which does not necessarily improve sampling performance.

\subsection{Time-intensive Sampling in Mini-batch Inference}

\begin{figure}[!ht] 
	\centering
	\vspace{-5pt}
	\includegraphics[width=0.48\textwidth]{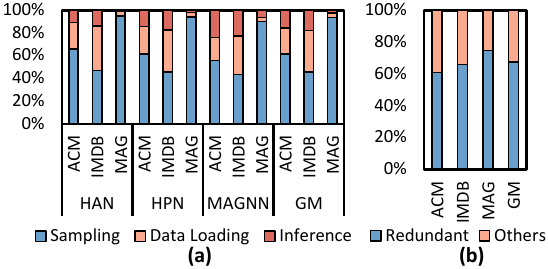}
\caption{Breakdown of mini-batch HGNN inference: (a) Time breakdown by phase; (b) Measured prefix-overlap ratio of reuse-eligible sampled path instances.}
        \vspace{-10pt}
	\label{fig:motivation_breakdown}
\end{figure}

As shown in Fig.~\ref{fig:motivation_breakdown}(a), we conduct experiments breaking down the execution time of each phase in mini-batch HGNN inference across several prominent HGNN models and datasets. The results demonstrate that, in end-to-end mini-batch HGNN inference, mini-batch sampling accounts for an average (geometric mean, GM) of 66.85\% of the total execution time, whereas the inference process itself contributes only 8.87\% on average. On the largest dataset OGBN-MAG (MAG), the sampling phase even accounts for more than 90\% of the total execution time. This significant time disparity emphasizes that mini-batch sampling constitutes the key performance bottleneck, underscoring the necessity to optimize this phase to accelerate the overall end-to-end mini-batch inference process.

\subsection{Memory-bound Characteristic of Sampling Phase}

The execution behavior of the sampling phase is fundamentally a graph traversal process, wherein the algorithm iteratively explores the neighborhood of target vertices and subsequently the neighborhoods of their neighbors. This process is characterized by extensive irregular memory access and low computational density, resulting in a memory-bound characteristic in nature. As presented in Table~\ref{tab:sampling_profile}, the Last Level Cache (LLC) hit rate during the sampling phase remains consistently low across various datasets, averaging only 34.03\%, reflecting the direct impact of irregular memory access patterns. Moreover, the negligible operation-to-byte ratios for both integer and floating-point operations indicate that the sampling phase is highly memory-bound with minimal computation. Floating-point operations are primarily used for transition-probability computation, while most integer instructions perform non-arithmetic operations, such as identifier comparisons and index computations. These findings indicate that the sampling phase is fundamentally memory-bound, with memory access being the principal performance bottleneck, rather than computational tasks.

\begin{table}[!th]
    \caption{Profiling results of mini-batch sampling across different datasets.}
    \label{tab:sampling_profile}
    \centering
    \setlength\tabcolsep{1.5pt}%
    \resizebox{0.49\textwidth}{!}
    {

    \begin{tabular}{cccc}
    \toprule
    \textbf{Dataset} & \textbf{LLC Hit Rate (\%)} & \textbf{Integer Ops./Byte} & \textbf{FLOPs/Byte} \\ \midrule \midrule
    ACM  & 41.23 & 0 & 0.0027 \\ 
    IMDB & 39.71 & 0 & 0.0011 \\ 
    MAG  & 24.08 & 0 & 0.0008 \\ \bottomrule
    \end{tabular}

}
\vspace{-12pt}
\end{table}

\subsection{Acceleration Opportunity from Semantic Redundancy}

As introduced by Section~\ref{sec:background_sampling}, unlike traditional GNN sampling, in metapath-based HGNN sampling process, each neighbor is obtained through traversal paths that involve various combinations of adjacency relations. Induced by semantic redundancy between longer and shorter metapaths, an inclusion relationship exists among the sampling traversal paths as in Fig.~\ref{fig:HetG}(a). We quantify this redundancy using the prefix-overlap ratio of reuse-eligible sampled path instances. For each longer metapath, we compare the prefixes of its sampled path instances with the independently sampled path instances of its longest reusable prefix. The prefix-overlap ratio is the percentage of longer-path instances whose prefixes exactly match the corresponding sampled path instances of the reusable prefix. Considering only the longest reusable prefix ensures that each reusable path instance is counted only once. Fig.~\ref{fig:motivation_breakdown}(b) shows a geometric mean prefix-overlap ratio of 66.93\% across ACM, IMDB, and MAG. This finding highlights the prevalence of semantic redundancy and identifies a significant opportunity to mitigate it, thereby expediting the sampling process and enhancing the overall performance of end-to-end mini-batch HGNN inference.

\subsection{Challenges Faced by Traditional Platforms}


Both CPUs and GPUs are inefficient for HGNN sampling due to the irregular memory access patterns inherent in this process. CPUs, optimized for sequential tasks, struggle with the non-linear memory access required, leading to frequent cache misses and high memory latency. While GPUs excel in parallel processing for tasks like matrix multiplication, they perform poorly in HGNN sampling, which involves traversing complex, multi-semantic paths with varying relations, resulting in poor memory locality and underutilization of parallelism. Thus, both architectures face significant bottlenecks in handling the non-sequential, high-memory-demand nature of HGNN sampling, making them suboptimal for this task.

\section{Design}

This section first provides an overview of ESR-HGNN, followed by a detailed introduction to the proposed redundancy-aware HGNN sampling paradigm designed to reduce irregular memory accesses. Subsequently, we present the architecture of the corresponding sampling unit and propose an end-to-end pipelined scheduling method to maximize execution parallelism within a single processing channel. Finally, we introduce a reusability-driven semantic grouping strategy to further enhance the efficacy of the redundancy-aware sampling approach in scenarios that leverage semantic parallelism.

\subsection{Overview}

In Section~\ref{sec:motivation}, we identify the sampling phase as the primary performance bottleneck in end-to-end mini-batch HGNN inference, characterized by a pronounced memory-bound nature and substantial semantic redundancy. ESR-HGNN addresses these challenges through three coordinated optimizations. First, redundancy-aware sampling eliminates redundant DRAM accesses by reusing the sampled results of matched metapath prefixes. Second, a specialized sampling unit and optimized pipeline overlap trie management with sampling to reduce exposed online overhead. Third, reusability-driven semantic grouping maximizes the effectiveness of prefix reuse when semantic parallelism is available.

\begin{figure*}[!t]
\centering
\includegraphics[width=0.98\textwidth]{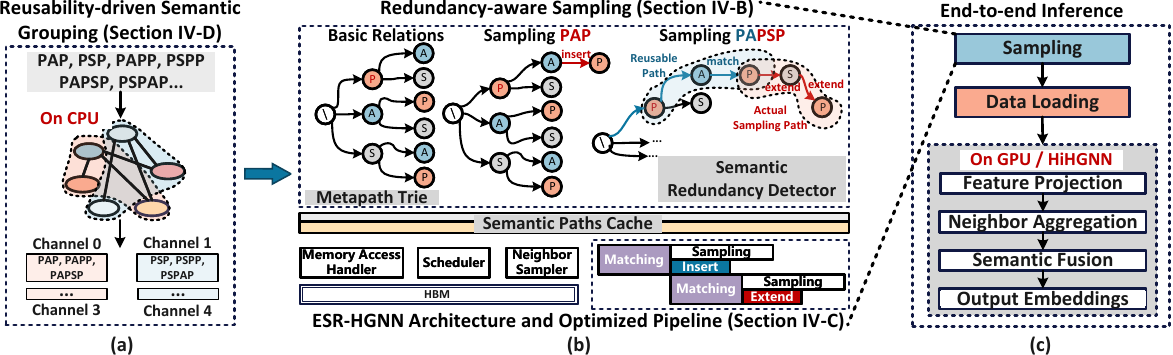}
\vspace{-5pt}
\caption{End-to-end workflow of ESR-HGNN. (a) One-time reusability-driven semantic grouping and channel assignment. (b) Redundancy-aware sampling with PAP-prefix reuse for PAPSP, together with the specialized sampling architecture and intra-task pipeline overlap. (c) Transfer of sampled mini-batches for HGNN inference on a GPU or HiHGNN.}
\label{fig:workflow}
\vspace{-8pt}
\end{figure*}

Fig.~\ref{fig:workflow} presents the end-to-end workflow of ESR-HGNN. Its core mechanism is the redundancy-aware sampling paradigm shown in Fig.~\ref{fig:workflow}(b), which leverages a trie to identify shared metapath prefixes. Using the PAP and PAPSP example, PAP is first inserted and sampled. PAPSP then matches the PAP prefix, retrieves the reusable result from the \textit{Semantic Paths Cache}, and samples only the remaining suffix. The lower part of this panel illustrates the specialized sampling architecture and the pipeline overlap between trie management and sampling. In the metapath-based setting considered here, sampling tasks for different metapaths are data-independent, enabling semantic parallelism. To exploit this property while preserving the effectiveness of redundancy-aware sampling, reusability-driven semantic grouping is performed once on the host CPU as a preprocessing step as shown in Fig.~\ref{fig:workflow}(a). The resulting groups are then assigned to parallel sampling channels. Finally, Fig.~\ref{fig:workflow}(c) shows that the sampled mini-batches are transferred to a GPU or HiHGNN for data loading and HGNN inference.

\subsection{Redundancy-aware Sampling}





In metapath-based HGNN sampling, the semantic redundancy between multiple metapaths directly leads to massive redundant traversal paths, thereby limiting the performance of the sampling process. To efficiently reuse the traversal paths sampled from shorter metapaths, we propose a redundancy-aware sampling paradigm, as shown in Algorithm~\ref{alg:metapath_sampling}. It records sampled metapaths in a Metapath Trie and reuses the sampled result of a matched prefix, so that only the unmatched suffix of a longer metapath is traversed. For each target vertex, the metapaths are processed in nondecreasing length order, ensuring that a reusable shorter prefix is available before a longer metapath is considered.






\begin{algorithm}[!t]
\SetAlgoLined
\SetKwBlock{DoParallel}{concurrently for $(v,m)$}{end}
\caption{Redundancy-aware Sampling Paradigm}
\label{alg:metapath_sampling}
\SetKwInOut{Input}{Input}
\SetKwInOut{Output}{Output}
\SetKwInOut{Initial}{Initial}

\Input{Heterogeneous graph $G=(V,E)$, \\
Target vertices batch $V_B$, \\
Metapaths $M = (m_1, m_2, \ldots, m_k)$, \\
Metapath Trie $T$, \\
Number of sampled neighbors $n$}
\Output{Mini-batch $g = \{g_{m_1} \cup g_{m_2} \cup \ldots \cup g_{m_k}\}$}
\Initial{$T\ \leftarrow \emptyset$, $(g_{mi}\ for\ g_{mi}\ in\ g) \leftarrow \emptyset$}

\For{each target vertices $v \in V_B$}{

    \For{each $m \in M$}{ \Comment{\textit{vertex-wise sampling}} \\
        $N_v^m\ \leftarrow \{v\}$  \\
        \If{$m$ has a reusable prefix $m'$ in $T$}{  \Comment{\textit{trie-based redundancy detection}} \\
            \DoParallel{
                $T \leftarrow \text{ExtendTrie}(T,m',m)$ \\
                \For{each neighbor $v_n \in N_v^{m'}$}{
                    $N_v^m \leftarrow \text{Sample}(G, v_n, m-m', 1)$ \\
                }
            }
        }
        \Else{
            \DoParallel{
                $T \leftarrow \text{Insert}(m)$ \\
                $N_v^m \leftarrow \text{Sample}(G, v, m, n)$ \\
            }
        }
        $g_m$ $\leftarrow$ $N_v^m$, $g_m$ $\leftarrow$ $\{e | e_{(v,N_v^m)} \in E\}$)
    }
}
\Return $g$

\end{algorithm}

Algorithm~\ref{alg:metapath_sampling} adopts a vertex-wise traversal order in the outer loop (lines~1-22). For each target vertex $v$, the inner loop iterates over every metapath $m$ and initializes its sampled neighborhood $N_v^m$ (lines~2-4). The Metapath Trie encodes each metapath as a root-to-leaf sequence of typed trie nodes, enabling longest-prefix matching. The trie representation and hardware implementation are described in Section~\ref{sec:architecture}. If the current metapath $m$ matches a reusable prefix $m'$ in the trie (line~5), the algorithm reuses the corresponding sampled result and transforms the original sampling task into new one. Using the PAP and PAPSP example in Fig.~\ref{fig:workflow}(b), PAP is first inserted and sampled. PAPSP then matches PAP, retrieves $N_v^{m'}$ from the Semantic Paths Cache, and extends each reused endpoint by sampling the residual suffix $m-m'$ (lines~9-11), thereby avoiding resampling the matched prefix. \textit{ExtendTrie} in line~8 extends, rather than replaces $m'$. It preserves child trie nodes with matching type codes, allocates new nodes only for missing suffix segments, and marks the terminal node of $m$ with \textit{end\_flag} while preserving the terminal marker of $m'$. If no reusable prefix is found, lines~14-19 insert $m$ into the trie and sample the complete metapath from $v$. Finally, line~20 adds the sampled vertices and their associated edges to the corresponding metapath-specific mini-batch. Note that the concurrent blocks in lines~7-12 and lines~15-18 operate only within a fixed target-metapath pair $(v,m)$. They overlap the independent trie-management operation (\textit{ExtendTrie} in line~8 or \textit{Insert} in line~16) with the corresponding neighbor sampling, rather than parallelizing iterations over target vertices or metapaths.

Our sampling paradigm differs from traditional approaches in two key aspects. \textit{\underline{First}}, we employ a metapath trie to record previously sampled metapaths and dynamically adjust sampling tasks based on matching results. This mechanism enables the direct reuse of previously traversed paths, thereby eliminating a substantial number of unnecessary irregular memory accesses. \textit{\underline{Second}}, traditional sampling paradigms prioritize metapath-wise processing for the simplicity of workflow, performing sampling for all target vertices within each metapath. Our approach instead utilizes a vertex-wise manner, which sequentially samples the same target vertex across multiple metapaths. This design offers two primary advantages. On one hand, it reduces the storage overhead for sampling paths of shorter prefix metapaths, as the cached sampling sequences for a given target vertex can be quickly reused when sampling other metapaths for the same vertex, eliminating the need for further storage. On the other hand, it allows the adjacency information of the same target vertex across different relations to remain on-chip for an extended period, thereby further reducing off-chip memory accesses, as the adjacency data is reused multiple times for different metapaths.

This method targets the standard metapath-guided neighbor-sampling workflow, in which sampling paths are independently extended according to their predefined metapaths. Prefix reuse does not change the per-hop sampling rule for the residual suffix. In this work, we instantiate the design with the \textit{RandomWalk} sampler~\cite{randomwalk}, following the metapath-guided random-walk paradigm of Metapath2Vec~\cite{metapath2vec}.



\subsection{Architecture of ESR-HGNN}
\label{sec:architecture}

\begin{figure*}[!ht] 
	\centering
	\vspace{-5pt}
	\includegraphics[width=0.96\textwidth]{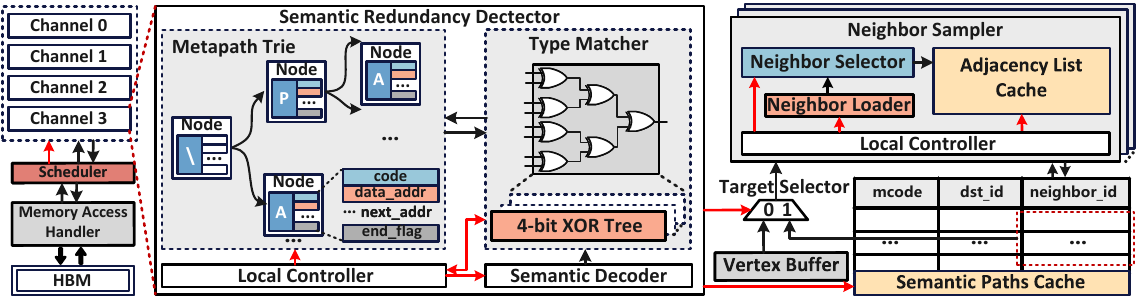}
\caption{Architecture of ESR-HGNN. }
        \vspace{-15pt}
	\label{fig:architecture}
\end{figure*}

To efficiently map the redundancy-aware sampling method, we design a novel specialized HGNN sampling unit named ESR-HGNN, as illustrated in Fig.~\ref{fig:architecture}. We adopt a multi-channel design to leverage the inherent parallelism between different semantics, as the sampling processes for different metapaths are entirely independent and free from data dependencies.


\subsubsection{Hardware Components}

The hardware architecture of ESR-HGNN is divided into three main components, including the \textit{Semantic Redundancy Detector}, the \textit{Neighbor Sampler} and the storage component. 

The \textit{Semantic Redundancy Detector} mainly includes a metapath trie for recording previously sampled metapaths. A trie is a tree-like data structure that efficiently stores and retrieves sequences like strings, with each node representing a character and paths from the root to the leaves corresponding to different strings or prefixes. It performs prefix matching for subsequent metapaths, and if a complete path is matched, the traversal paths of the matched metapath are directly reused. To enable efficient prefix matching, we utilize binary encoding for metapaths and design a matching unit based on a 4-bit XOR tree, as 4-bit encoding is generally sufficient for typical HetGs. This design facilitates metapath node matching with exceptionally low latency and energy consumption, and it can be modularly scaled to accommodate longer metapath encodings, supporting broader matching operations. Unlike linear comparison and hash-based lookup, the trie natively represents shared prefixes and identifies the longest reusable prefix without sequential comparisons or repeated prefix lookups. Although CAM-based matching supports parallel lookup, replicating its storage and comparison logic across sampling channels incurs significantly higher hardware cost. Therefore, the trie-based design offers a more favorable functionality-cost tradeoff for the short metapaths considered in this work.

Specifically, each channel stores its trie in private on-chip memory as fixed-format node records. The \textit{next\_addr} vector holds the addresses of valid child records, with invalid entries denoting absent children. During traversal, the \textit{Local Controller} follows this vector to fetch the candidate child records, and the \textit{Semantic Decoder} sends the requested vertex-type code to the XOR-based \textit{Type Matcher}. The matcher compares this code with the \textit{code} field of each candidate and returns the matching child address. Therefore, trie traversal is determined by explicit links and type codes, rather than by the physical placement order of node records. After a complete metapath is matched, its \textit{data\_addr} descriptor which is not involved in type matching identifies the corresponding sampled-result block for retrieval from the \textit{Semantic Paths Cache}.

Only completed metapaths whose terminal trie nodes are marked with \textit{end\_flag} generate reusable result blocks; intermediate trie nodes do not occupy the Semantic Paths Cache. When the cache is full, FIFO replacement writes evicted blocks back to HBM and reloads them on demand. Because metapaths are processed in nondecreasing length order, this policy naturally retains recently sampled, typically longer prefixes that are more likely to be reused by subsequent longest-prefix matches.

Another main component is the \textit{Neighbor Sampler}, which traverses neighbors in compressed sparse row format for target vertex and repeats this process according to the semantic sequence until a sufficient number of neighbors are sampled as required. It mainly consists of a \textit{Neighbor Loader} responsible for loading the neighbor list of the current target vertex from memory, and a \textit{Neighbor Selector} constituted by linear-feedback shift registers (LFSR) that randomly select neighboring vertices. The private \textit{Adjacency List Cache} is used to store the vertex's neighbors for reuse. It is a frequency-aware cache that retains adjacency data of frequently accessed vertices for longer. Since it cannot hold the entire graph, evicted or low-frequency adjacency data are fetched from HBM on demand. Note that each channel contains multiple basic \textit{Neighbor Sampler}s to support the parallel sampling of different target vertices or different metapaths for the same target vertex.

The storage component is primarily composed of a \textit{Vertex Buffer} and a \textit{Semantic Path Cache}. The \textit{Vertex Buffer} is chiefly responsible for pre-caching the target vertices within the batch to be processed. The primary function of the \textit{Semantic Path Cache} is to store traversal paths derived from the sampled neighbors of various vertex types, facilitating the replacement of the original target vertex with a new one upon the identification of redundancy. Specifically, when the \textit{Semantic Redundancy Detector} identifies that the current metapath being processed has a reusable sub-metapath that has already been sampled through metapath matching, the \textit{Semantic Redundancy Detector} sets the control signal of the target selector to 1, and directly retrieves the neighbors sampled based on the matched sub-metapath from the \textit{Semantic Paths Cache}, using them as the new starting points for sampling thus eliminating redundant traversals. For the metapaths remaining after prefix matching (metapath $m-m'$ in line 10 of Algorithm~\ref{alg:metapath_sampling}), no further prefix matching is performed. Instead, matching is based solely on the target vertex and the metapath itself in the cache, and reuse occurs only if the \textit{Semantic Path Cache} contains the corresponding entry.



\subsubsection{Optimized Pipeline}

Building on the aforementioned hardware components, we present a fine-grained pipeline dataflow to maximize execution parallelism within a single channel in end-to-end mini-batch HGNN inference.

\begin{figure}[!htb] 
	\centering
	\includegraphics[width=0.48\textwidth]{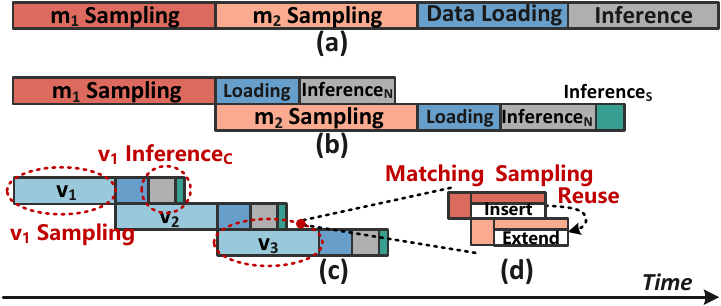}
\caption{End-to-End pipeline: (a) Original execution paradigm; (b) Naive pipeline; (c) Optimized pipeline; (d) intra-task overlap between trie management and sampling for a fixed target-metapath pair $(v,m)$.}
        \vspace{-5pt}
	\label{fig:end2end_pipeline}
\end{figure}

Consider an end-to-end mini-batch inference process with two metapaths $m_1, m_2$ and three target vertices $v_1, v_2, v_3$, as illustrated in Fig.~\ref{fig:end2end_pipeline}(a), on traditional platforms, HGNN mini-batch inference is typically executed in a sequential manner, where the sampling phase precedes the loading and inference phase. During sampling, a metapath-wise sampling paradigm is employed. Although this approach is straightforward to implement and does not introduce additional scheduling overhead, it lacks the ability to leverage parallelism across multiple dimensions, leading to suboptimal execution performance.

A straightforward optimization approach is to pipeline the sampling and inference processes for each metapath, allowing the sampling process of the next metapath to execute in parallel with the inference process of the previous metapath, as illustrated in Fig.~\ref{fig:end2end_pipeline}(b). However, this method has two key limitations. First, this execution order fails to exploit the available parallelism among sampling tasks associated with different metapaths. Second, after sampling the neighborhood of only one metapath, the resulting input remains semantically incomplete because representations from the remaining semantic graphs are unavailable. Consequently, semantic fusion ($Inference_S$ in Fig.~\ref{fig:end2end_pipeline}(b)) cannot be initiated or overlapped with the sampling and per-semantic-graph inference of the remaining metapaths. For each target vertex, $Inference_S$ can start only after the per-semantic-graph inference stage ($Inference_N$ in Fig.~\ref{fig:end2end_pipeline}(b)) has been completed for all metapaths.

Leveraging our novel sampling paradigm and specialized structural design, we realize a finer-grained parallelization approach in load scheduling from two key perspectives. \textit{\underline{First}}, the sampling process for the same target vertex across different metapaths is structured as a pipelined execution, which is divided into two distinct stages: prefix metapath matching and the actual sampling process, as shown in Fig.\ref{fig:end2end_pipeline} (d). Moreover, both the sampling process and the insertion and updating procedures of the metapath trie are fully parallelizable as lines 7-12 and 15-18 in Algorithm~\ref{alg:metapath_sampling}. This parallelism is an intra-task overlap between trie management and sampling for a fixed $(v,m)$ instead of parallel iterations over target vertices or metapaths. Therefore, the matching, insertion, and extending processes of the metapath trie required for detecting and eliminating semantic redundancy can be effectively overlapped with other execution processes, significantly reducing the associated overhead. \textit{\underline{Second}}, our vertex-wise sampling paradigm completes the neighborhoods of all metapaths for a target vertex before its inference is launched. The resulting semantically complete input enables $Inference_C$, including neighbor aggregation and semantic fusion, and thereby allows the semantic fusion stage to be parallelized. This fine-grained pipelined execution maximizes parallelism across various execution stages within a single channel, thereby substantially enhancing the overall end-to-end mini-batch inference performance.

\subsection{Reusability-driven Semantic Grouping}


The sampling tasks associated with different metapaths exhibit semantic parallelism because they have no data dependencies. To further optimize mini-batch sampling performance, the proposed redundancy-aware paradigm must support this semantic parallelism across hardware channels. However, metapaths must be grouped according to the number of available channels. Random grouping often reduces reusable semantic paths within a single channel, limiting the efficacy of the redundancy-aware paradigm and overall performance. To address this challenge, we propose a reusability-driven semantic grouping method that serves as data preprocessing to enhance redundancy within each group, maximizing performance in parallel settings.

\begin{figure}[!t] 
	\centering
	\includegraphics[width=0.48\textwidth]{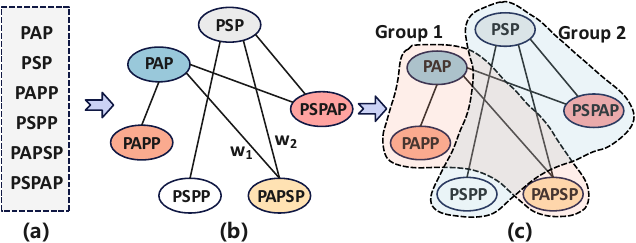}
\caption{A toy example of metapath grouping: (a) original metapaths; (b) weighted metapath-reuse graph; (c) resulting metapath groups.}
        \vspace{-15pt}
	\label{fig:metapath_grouping}
\end{figure}

Given a set of metapaths $M$, as illustrated in Fig.~\ref{fig:metapath_grouping}(a), we model each metapath as a vertex $v \in V_M$. An edge $e \in E_M$ is established between two vertices if their corresponding metapaths exhibit a reuse relationship. The resulting weighted metapath-reuse graph $G_M=(V_M,E_M)$, shown in Fig.~\ref{fig:metapath_grouping}(b), explicitly captures the reusability among metapaths. Moreover, the degree of reusability varies among prefix metapaths for the same target metapath. For example, in the case of PAPSP, the prefix metapath PAP demonstrates higher reusability than PSP, as both PAP and PAPSP initiate sampling from the same target vertices. However, when PAPSP extends its traversal to PSP, the starting vertices for PSP-based traversal are mostly no longer the original target vertices. To quantify this, we assign a weight to each edge, representing its degree of reusability. It's calculated as $w=1-id_{sm} / l_{lm}$, where $id_{sm}$ is the starting position where the shorter metapath aligns within the longer metapath, and $l_{lm}$ is the length of the longer target metapath. This metric captures the reusability degree of metapath vertices within the hypergraph, with higher weights indicating greater reusability. Based on this definition, in the given example, the weight $w_1$ is greater than $w_2$, signifying that PAP has higher reusability with respect to PAPSP than PSP.

Building on the constructed weighted metapath-reuse graph and the defined reusability weight metric, the problem of optimal metapath grouping is formulated as a community detection task within this graph, as illustrated in Fig.~\ref{fig:metapath_grouping}(c). The objective is to maximize the total sum of edge weights within each detected group. Inspired by the classical \textit{Louvain} algorithm~\cite{louvain}, we present a reusability-driven semantic grouping method as shown in Algorithm~\ref{alg:metapath_grouping}, incorporating the proposed reusability weight metric into the calculation of group reusability.

\begin{algorithm}[!t]
\SetAlgoLined
\caption{Reusability-driven Semantic Grouping}
\label{alg:metapath_grouping}
\SetKwInOut{Input}{Input}
\SetKwInOut{Output}{Output}
\SetKwInOut{Initial}{Initial}

\Input{Weighted metapath-reuse graph $G_M = (V_M, E_M)$, number of hardware channels $n$}
\Output{At most $n$ metapath groups}
\Initial{
    Initialize groups $C \gets \{v\ for\ v\ in\ V_M\}$  \\
    Set $improved \gets \textbf{True}$ \\
    Set $num\_groups \gets |C|$  \\
}

\centerline{{\underline{$\triangleleft \quad \textbf{Phase 1: Reusability-improving Step}$}}}

\While{$improved$ is \textbf{True}}{
    Set $improved \gets \textbf{False}$ \\
    \For{each vertex $v \in V_M$}{
        $\Delta Q_{max} \gets 0$, $C_t \gets \phi$ \\
        \For{each neighbor group $C'$ of $v$}{
            $\Delta Q \gets \text{ComputeReusabilityGain}(v, C')$ \\
            \If{$\Delta Q\ > \Delta Q_{max}$}{
                $\Delta Q_{max} \gets \Delta Q$, $C_t \gets C'$ \\
            }
        }
        \If{$\Delta Q_{max} > 0$}{
            Move $v$ to $C_t$, Set $improved \gets \textbf{True}$ \\
        }
    }
}

\centerline{{\underline{$\triangleleft \quad \textbf{Phase 2: Group-merging Step}$}}}

\While{$num\_groups > n$}{
    $G'_M \gets \text{BuildGraph}(C)$ \\
    Reapply \textbf{Phase 1} to $G'_M$ \\
    Set $num\_groups \gets |C|$ \\
}

\Return $C$
\end{algorithm}

It consists of two main parts: the \textit{Reusability-improving} and the \textit{Group-merging} steps. In the reusability-improving step, for each metapath vertex, the algorithm attempts to add it to each neighboring group and calculates the reusability gain for each (line 6). It identifies the neighbor with the highest reusability gain (lines 7-9), and if this maximum gain is greater than 0, the vertex is added to that neighbor's group (lines 11-13); otherwise, it remains in its original group. In the group-merging stage, further merging is required based on the maximum number of channels supported by the hardware. Specifically, if the number of groups generated in Phase 1 exceeds the number of available hardware channels, the groups are reconstructed into a new weighted metapath-reuse graph, and Phase 1 is repeated until the number of groups fits within the available channels (lines 16–20).

This strategy facilitates metapath grouping to maximize reusable paths within each group while exploiting the inherent semantic parallelism of metapath-based sampling. By colocating metapaths with shared prefixes, it increases prefix reuse and reduces redundant sampling and off-chip memory accesses, thereby enabling the redundancy-aware sampling method to fully realize its performance potential through parallel execution. The grouping algorithm is executed only once on the host before mini-batch inference, and its cost is amortized across all subsequent mini-batches, as quantified in Section~\ref{sec:overhead_analysis}. Community partitioning may result in imbalanced channel loads because the grouping objective explicitly prioritizes prefix reuse over metapath-count balance. However, enforcing strict load balance may distribute metapaths with shared prefixes across different channels, preventing sampled results from being reused and consequently increasing redundant HBM traffic. For the targeted memory-bound sampling workload, reducing off-chip memory traffic and the resulting post-reuse critical-path latency is more beneficial than equalizing nominal channel loads. Therefore, ESR-HGNN prioritizes prefix reuse over strict load balance to maximize overall sampling performance, while accepting moderate channel-load imbalance as a favorable trade-off.

\section{Evaluation}

In this section, we first describe the experimental setup and present a comprehensive evaluation of the overall performance. We then validate the effectiveness of the proposed optimizations through detailed ablation studies. Finally, we evaluate the impact of ESR-HGNN on inference accuracy, quantify the overhead introduced by each optimization, and investigate its sensitivity and scalability.

\subsection{Experiment Setup}
\label{sec:experiment_setup}

\textit{\textbf{Methodology.}} The performance and energy efficiency of ESR-HGNN are assessed utilizing the following tools.

\textit{Cycle-accurate Simulator.} 
We implement ESR-HGNN within a cycle-accurate simulator to assess its performance in terms of execution cycles, and integrate Ramulator~\cite{ramulator} to precisely model off-chip memory accesses to HBM.


\textit{CAD Tools.}
The Synopsys Design Compiler with the TSMC 12$nm$ standard VT library is employed for synthesizing the RTL implementation of each module. Power consumption is estimated using Synopsys PrimeTime PX. The module with the longest critical path delay measures 0.84 ns, enabling ESR-HGNN to reliably operate at a 1.0 GHz clock frequency.

\textit{Memory Measurements.}
The access latency, energy consumption, and area of on-chip memory components are estimated using Cacti 6.5~\cite{CACTI}. Four distinct scaling factors are applied to adjust these estimates to the 12nm technology node, following the approach in work~\cite{technology_scale}. The latency and energy consumption of HBM1.0 are simulated via Ramulator and estimated at 7 pJ/bit as in work~\cite{7pj}.

\textit{\textbf{Benchmarks.}}
Experiments are conducted on three prominent HGNN models, including HAN~\cite{HAN}, HPN~\cite{HPN}, and MAGNN~\cite{HetGNN}, implemented using the DGL 1.0.2 framework~\cite{DGL}. We use the ACM, IMDB, and MAG datasets provided by the OpenHGNN toolkit~\cite{openhgnn}, as detailed in Table~\ref{tab:datasets}. Among them, the largest dataset, MAG, contains edges at the billion scale under the given metapaths. The traditional \textit{RandomWalk}\cite{randomwalk} neighbor sampling method is utilized as the foundational sampling technique. Unless specified, for the ACM and IMDB datasets, the batch size for sampling is configured to 128, with the number of sampled neighbors set to 20. For the MAG dataset, the batch size is set to 1024, and the number of sampled neighbors is configured to 1000 considering its large scale.

\begin{table}[!ht]
\caption{Information of HetG datasets.}
\label{tab:datasets}
\resizebox{0.49\textwidth}{!}{
\begin{tabular}{cccc}
\toprule
\textbf{Dataset} &
  \textbf{\#Vertex} &
  \textbf{\#Edge of Relations} &
  \textbf{Metapaths} \\ \midrule \midrule
\multirow{4}{*}{ACM} &
  Paper: 3025 &
  \multirow{4}{*}{\begin{tabular}[c]{@{}c@{}}PP: 5343 -PP:5343\\ AP: 9936 PA: 9936\\ PS: 3025 SP: 3025\end{tabular}} &
  \multirow{4}{*}{\begin{tabular}[c]{@{}c@{}}PAP / PSP\\ PAPP / PSPP\\ PAPSP / PSPAP\end{tabular}} \\
 &
  Author: 5912 &
   &
   \\
 &
  Subject: 56 &
   &
   \\
 &
  Term:1902 &
   &
   \\ \midrule
\multirow{4}{*}{IMDB} &
  Movie: 4278 &
  \multirow{4}{*}{\begin{tabular}[c]{@{}c@{}}AM: 12828 DM: 4278\\ MA: 12828 MD: 4278\\ KM: 23610 MK: 23610\end{tabular}} &
  \multirow{4}{*}{\begin{tabular}[c]{@{}c@{}}MDM / MAM / MKM\\ MDMAM / MAMDM\\ MDMKM / MKMDM\\ MAMKM / MKMAM\end{tabular}} \\
 &
  Director:2081 &
   &
   \\
 &
  Actor:5257 &
   &
   \\
 &
  Keyword:7971 &
   &
   \\ \midrule
\multirow{4}{*}{MAG} &
  Author:1134649 &
  \multirow{4}{*}{\begin{tabular}[c]{@{}c@{}}PP: 5416271 -PP: 5416271\\ AI: 1043998 IA: 1043998\\ AP: 7145660 PA: 7145660\\ PF: 7505078 FP: 7505078\end{tabular}} &
  \multirow{4}{*}{\begin{tabular}[c]{@{}c@{}}PAP / PAPP\\ PFP / PAPFP\\ PFPAP / PAIAP\end{tabular}} \\
 &
  Paper: 736389 &
   &
   \\
 &
  Field: 59965 &
   &
   \\
 &
  Institute:8740 &
   &
   \\ \bottomrule
\end{tabular}
}
\end{table}

\textit{\textbf{Platforms.}}
Performance evaluations are carried out on an Intel\textsuperscript{\textregistered} Xeon\textsuperscript{\textregistered} Platinum 8350C CPU and an NVIDIA A100 GPU. CPU is used solely for sampling, while the GPU mainly serves as the inference platform, and is also used as a comparison baseline for sampling performance due to DGL's support for GPU-based sampling. The CPU-related performance metrics are obtained through Perf, Intel VTune and Intel Advisor, while the GPU-related performance metrics are collected using NVIDIA Nsight Compute, with all experiments conducted in Float32 precision. Additionally, the SOTA HGNN inference accelerator, HiHGNN~\cite{HiHGNN}, is also introduced as the baseline inference platform. Table~\ref{tb:platform} summarizes the hardware configurations used for the CPU and GPU baselines and for ESR-HGNN simulation.

\begin{table*}[!t]
\centering
\caption{Hardware configurations of the baseline platforms and ESR-HGNN.}
\label{tb:platform}
\renewcommand\arraystretch{1.12}
\setlength{\tabcolsep}{10pt}
\begingroup
\begin{tabular}{lccc}
\toprule
 & \textbf{CPU} & \textbf{GPU} & \textbf{ESR-HGNN} \\ \midrule
Compute resources & 32 cores / 64 threads & \begin{tabular}[c]{@{}c@{}}108 SMs / 6912 CUDA cores\\221,184 max. concurrent threads\end{tabular} & 4 sampling channels \\
CPU sampling workers & 32 (one per physical core) & -- & -- \\
Clock frequency & 2.6 GHz & 1.4 GHz & 1.0 GHz \\ \midrule
On-chip caches (total) & \begin{tabular}[c]{@{}c@{}}L1: 1.5 MB; L2: 40 MB\\L3: 48 MB (shared)\end{tabular} & \begin{tabular}[c]{@{}c@{}}L1: 20.25 MB; L2: 40 MB\\(shared)\end{tabular} & \begin{tabular}[c]{@{}c@{}}Semantic Paths Cache: 4 MB\\Adjacency List Cache: 2 MB\end{tabular} \\
Off-chip memory & 8-channel DDR4-3200 & 80 GB HBM2e & 2 GB HBM1.0 \\
Peak memory bandwidth & 204.8 GB/s (theoretical) & 2039 GB/s & 512 GB/s \\
Host interface & -- & \multicolumn{2}{c}{PCIe 4.0 $\times$16, 64 GB/s} \\ \bottomrule
\end{tabular}
\endgroup
\vspace{-6pt}
\end{table*}


\subsection{Overall Results}

\subsubsection{Speedup}

\begin{figure*}[!ht] 
	\centering
	\vspace{-5pt}
	\includegraphics[width=0.98\textwidth]{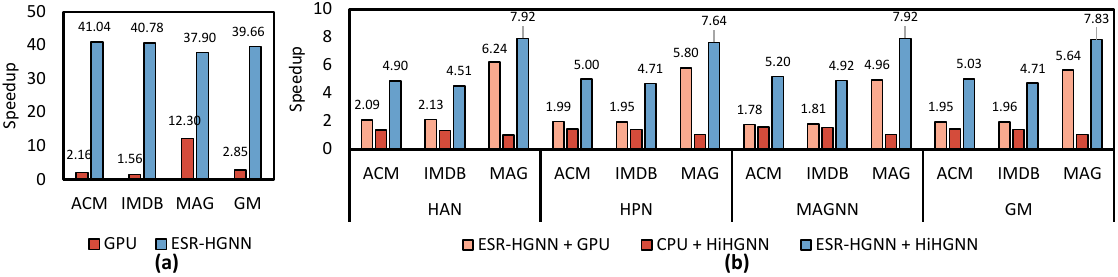}
	\caption{Speedup: (a) Pure sampling process over CPU; (b) End-to-end mini-batch HGNN inference over CPU+GPU.}
        \vspace{-10pt}
	\label{fig:speed_up}
\end{figure*}

As illustrated in Fig.~\ref{fig:speed_up}(a), ESR-HGNN achieves an average speedup of 39.66$\times$ over the CPU and 13.94$\times$ over the GPU in terms of metapath-based sampling. Furthermore, when integrated with the SOTA HGNN inference accelerator HiHGNN, ESR-HGNN+HiHGNN delivers an average end-to-end mini-batch inference speedup of 5.70$\times$ compared with the CPU+GPU setup, as shown in Fig.~\ref{fig:speed_up}(b). When combined with a GPU, ESR-HGNN+GPU achieves an average speedup of 2.78$\times$. Notably, CPU+HiHGNN delivers only a 1.15$\times$ average speedup in end-to-end performance over the CPU+GPU setup, highlighting that solely accelerating the inference phase is insufficient to substantially improve overall mini-batch inference performance.

ESR-HGNN's performance gains in sampling stem from three main factors. First, the novel redundancy-aware sampling paradigm significantly reduces redundant traversals, thereby minimizing unnecessary irregular memory accesses. Second, the specialized hardware design and optimized data flow eliminate redundant operations typical of general-purpose platforms and boost the parallelism between execution phases. Finally, the reusability-driven grouping method exploits semantic parallelism via multi-channel support and maximizes path reuse within individual channels. The end-to-end acceleration also comes from the optimized pipeline, which reduces the overhead by overlapping the data loading process.



\subsubsection{Reduction of DRAM Access and Energy Consumption}

\begin{figure}[!t] 
	\centering
	\includegraphics[width=0.48\textwidth]{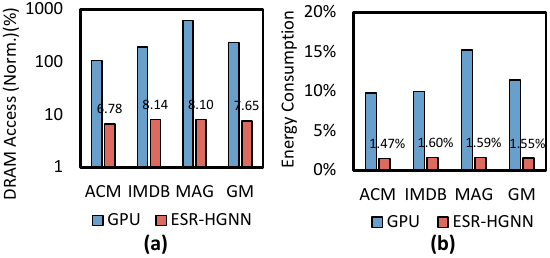}
	\caption{Results of sampling process normalized to CPU: (a) DRAM access; (b) Energy consumption.}
	\label{fig:dram_energy}
\end{figure}

As depicted in Fig.~\ref{fig:dram_energy}(a), ESR-HGNN achieves average DRAM access savings of 92.35\% and 96.75\% compared with CPU and GPU, respectively. The reduction in memory access is attributed to our proposed redundancy-aware sampling method as well as the fine-grained memory access brought about by the customized hardware structure and data path. Fig.~\ref{fig:dram_energy}(b) illustrates that ESR-HGNN reduces energy consumption by 98.45\% and 86.41\% compared with CPU and GPU, respectively. The reduction in energy consumption, in addition to the decrease in memory access, also results from the structural design that eliminates the overhead of general-purpose platforms, such as maintaining the operating system and software programming frameworks.

\subsubsection{Area and Power}

As presented in Table~\ref{tb:chip_area_power}, ESR-HGNN incorporates 6.53 MB of on-chip SRAM, 128 4-bit XOR trees, 2048 LFSRs, and additional control logic, occupying a total area of 5.01 $mm^2$ and consuming 559.62 $mW$ of power. Its area and power consumption account for only 23.47\% and 4.66\%, respectively, of those of HiHGNN, while achieving significant speedup in end-to-end mini-batch HGNN inference.

\begin{table}[!t]
\vspace{-3pt}
 \caption{Characteristics of ESR-HGNN (TSMC 12 $nm$).} \label{tb:chip_area_power}
 \centering
 \renewcommand\arraystretch{1.0}
    \resizebox{0.49\textwidth}{!}{
\begin{tabular}{|l|l|r|r|r|r|}
\hline
\multicolumn{2}{|l|}{\begin{tabular}[c]{@{}c@{}}\textbf{Component or Block}\end{tabular}} & \begin{tabular}[r]{@{}c@{}}\textbf{Area ($mm^2$)}\end{tabular} & \textbf{\%} & \begin{tabular}[r]{@{}c@{}}\textbf{Power ($mW$)}\end{tabular}  & \textbf{\%} \\ \hline \hline
\multicolumn{2}{|l|}{{ESR-HGNN}}    &5.01 &100 &559.62 &100    \\ \hline \hline
\multicolumn{6}{|c|}{\begin{tabular}[c]{@{}c@{}} \textbf{Breakdown by Functional Block} \end{tabular}} \\ \hline
\multicolumn{2}{|l|}{Semantic Paths Cache}   &2.95 &58.80 &332.62 &59.44    \\
\multicolumn{2}{|l|}{Adjacency List Cache}   &1.47 &29.40 &166.31 &29.72     \\
\multicolumn{2}{|l|}{Vertex Buffer}   &0.37 &7.35 &41.58 &7.43    \\
\multicolumn{2}{|l|}{Metapath Trie}     &0.02 &0.46 &2.60 &0.46    \\
\multicolumn{2}{|l|}{LFSR Array}    &0.01 &0.14 &0.20 &0.04  \\
\multicolumn{2}{|l|}{Others}          &0.19 &3.85 & 16.31 & 2.91    \\  \hline
\end{tabular}
}
\vspace{-10pt}
\end{table}

\subsection{Effects of Optimizations}
\label{sec:ablation}

\begin{figure*}[!ht] 
	\centering
	\vspace{-7pt}
	\includegraphics[width=0.96\textwidth]{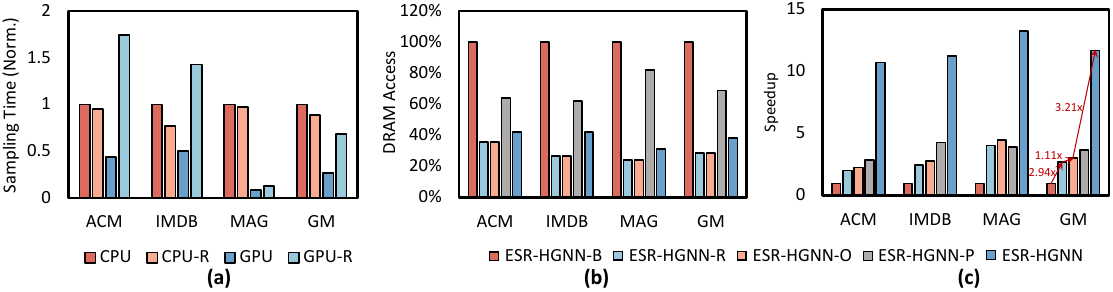}
	\caption{Effects of optimizations: (a) Software implementation of redundancy-aware sampling on CPU and GPU; (b) Number of DRAM access on ESR-HGNN; (c) Speedup on ESR-HGNN.}
        \vspace{-12pt}
	\label{fig:incremental_study}
\end{figure*}

Since our optimization approaches are hardware-software co-design solutions and cannot independently yield meaningful impact, we adopt a cumulative evaluation approach rather than an ablation study to demonstrate their effectiveness. Fig.~\ref{fig:incremental_study}(b) and (c) shows the impact of incrementally applying the optimizations. The \textbf{-B} configuration represents a baseline single-channel ESR-HGNN with naive sampling. The \textbf{-R} configuration introduces redundancy-aware sampling. The \textbf{-O} configuration adds the optimized pipeline. The \textbf{-P} configuration includes a 4-channel setup with random metapath grouping. The fully optimized configuration integrates reusability-based semantic grouping, demonstrating the cumulative performance improvements achieved.

\textbf{Effects of Redundancy-aware Sampling.} Since the redundancy-aware sampling paradigm can, in theory, be implemented in software, we evaluate its effectiveness by implementing it on both CPU and GPU platforms. As shown in Fig.\ref{fig:incremental_study}(a), the software-based approach yields limited benefits, achieving only a 1.12$\times$ speedup on the CPU, while actually increasing sampling time on the GPU due to diminished memory coalescing. In contrast, on the ESR-HGNN platform, the fine-grained memory access enabled by the customized hardware architecture and data path, along with the enhanced on-chip data locality provided by the dedicated cache, allows \textbf{-R} to significantly reduce DRAM access by an average of 71.56\% compared with \textbf{-B}, leading to a 2.94$\times$ improvement in sampling performance, as illustrated in Fig.\ref{fig:incremental_study}(b) and (c).

\textbf{Effects of Optimized Pipeline.} As shown in Fig.\ref{fig:incremental_study}(b), \textbf{-O} exhibits no significant change in memory access volume compared with \textbf{-R} since fine-grained pipelined execution preserves the original neighbor traversal behavior while enabling phase-overlapping execution. This optimization effectively parallelizes the management operations of the metapath trie with the sampling process, resulting in a 1.11$\times$ improvement in sampling performance, as illustrated in Fig.\ref{fig:incremental_study}(c).

\textbf{Effects of Reusability-driven Semantic Grouping.} While \textbf{-P} enables parallel sampling and has the potential to improve performance over \textbf{-O}, it reduces the number of reusable traversal paths within groups, diminishing the effectiveness of redundancy-aware sampling and yielding only marginal gains. On the MAG dataset, \textbf{-P} even leads to a 10.31\% performance drop, as shown in Fig.~\ref{fig:incremental_study}(c). In contrast, our reusability-driven grouping method enhances redundancy within groups, significantly reducing DRAM access and fully leveraging the benefits of redundancy-aware sampling. When combined with semantic parallelism, it achieves a 3.21$\times$ performance improvement over the non-parallel \textbf{-O} version.






\subsection{The Impact on Accuracy}

\begin{figure}[!ht] 
	\centering
	\vspace{-7pt}
	\includegraphics[width=0.48\textwidth]{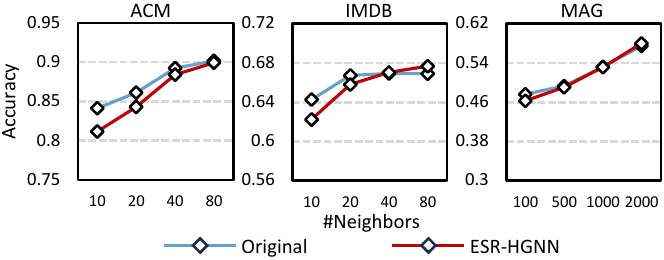}
	\caption{Averaged inference accuracy.}
	\label{fig:accuracy_loss}
	\vspace{-12pt}
\end{figure}

Reusing previously sampled traversal paths reduces partial traversal randomness, potentially affecting inference accuracy. Fig.\ref{fig:accuracy_loss} shows its impact across datasets and number of sampled neighbor settings on HAN model. On average, our method incurs only a 1.05\% accuracy drop compared with \textit{RandomWalk}~\cite{randomwalk}, a minor trade-off given the significant sampling efficiency gains. This minimal impact stems from two factors. First, prior studies~\cite{DropOut, DropNode} have demonstrated that GNNs exhibit robustness to partial information loss, a property that extends to HGNNs as well. Second, the reuse mechanism is restricted to prefix matching, ensuring that randomness is preserved in the subsequent sampling steps.

\subsection{Overhead Analysis}
\label{sec:overhead_analysis}

This section quantifies the online overhead of redundancy-aware sampling and the one-time preprocessing overhead of reusability-driven semantic grouping.

\textbf{Online Overhead of Redundancy-aware Sampling.}
Table~\ref{tab:sampling_overhead} reports the overhead relative to pure neighbor sampling, including adjacency list retrieval and random neighbor selection. \textit{Total online overhead} represents the latency of metapath trie insertion, matching, and extending, Semantic Paths Cache lookup and access, and scheduling. It accounts for 21.14\% of pure sampling latency on average across the evaluated datasets.

\begin{table}[!ht]
\vspace{-5pt}
\begingroup
\color{blue}
\caption{Online overhead of redundancy-aware sampling.}
\label{tab:sampling_overhead}
\endgroup
 \centering
 \renewcommand\arraystretch{1.0}
 {
    \resizebox{0.49\textwidth}{!}{
\begin{tabular}{|c|c|c|c|}
\hline
\textbf{Dataset} & \begin{tabular}[c]{@{}c@{}}\textbf{Pure Neighbor}\\\textbf{Sampling}\end{tabular} & \begin{tabular}[c]{@{}c@{}}\textbf{Total Online}\\\textbf{Overhead}\end{tabular} & \begin{tabular}[c]{@{}c@{}}\textbf{Visible Online}\\\textbf{Overhead}\end{tabular} \\ \hline
ACM  & 100\% & 18.01\% & 5.42\% \\ \hline
IMDB & 100\% & 21.0\% & 7.1\% \\ \hline
MAG  & 100\% & 25.0\% & 12.6\% \\ \hline
GM   & 100\% & 21.14\% & 7.86\% \\ \hline
\end{tabular}
}
}
\vspace{-8pt}
\end{table}

As introduced in Section~IV-C, the optimized pipeline overlaps online management operations with the post-matching sampling process. Consequently, the visible online overhead is reduced to only 7.86\% of pure sampling latency. This overhead is a worthwhile tradeoff for the substantial reduction in DRAM accesses achieved by redundancy-aware sampling paradigm.

Note that the overhead ratio is larger on MAG, since its larger graph, batch size, and sampling fanout, i.e., the number of sampled neighbors per target vertex, enlarge the working set. Under the fixed Semantic Paths Cache capacity, the cache hit rate decreases and the retrieval latency of matched sampling paths increases.

\textbf{One-time Overhead of Reusability-driven Semantic Grouping.}
We further assess the overhead of reusability-driven semantic grouping, which is a one-time preprocessing step on the host CPU. Table~\ref{tab:preprocessing_overhead} shows results for the IMDB dataset, which uses the most metapaths. On the ESR-HGNN+GPU platform, a single-batch inference is 29.52$\times$ the preprocessing time, and it is 12.29$\times$ on ESR-HGNN+HiHGNN. Because grouping is executed only once before all mini-batches are processed, its impact on total execution time is negligible while enabling significant performance gains.

\begin{table}[!ht]
\vspace{-5pt}
 \caption{Metapath Grouping overhead (normalized time).} 
 \label{tab:preprocessing_overhead}
 \centering
 \renewcommand\arraystretch{1.0}
    \resizebox{0.49\textwidth}{!}{
\begin{tabular}{|c|c|c|c|}
\hline
\textbf{Model} & \textbf{Metapath Grouping} & \textbf{+ GPU} & \textbf{+ HiHGNN} \\ \hline
HAN   & 1                 & 24.67  & 11.65     \\ \hline
HPN   & 1                 & 29.35  & 12.21     \\ \hline
MAGNN & 1                 & 35.51  & 13.32     \\ \hline
\end{tabular}
}
\vspace{-12pt}
\end{table}

\subsection{Sensitivity and Scalability Exploration}

\begin{figure}[!ht] 
	\centering
	\includegraphics[width=0.48\textwidth]{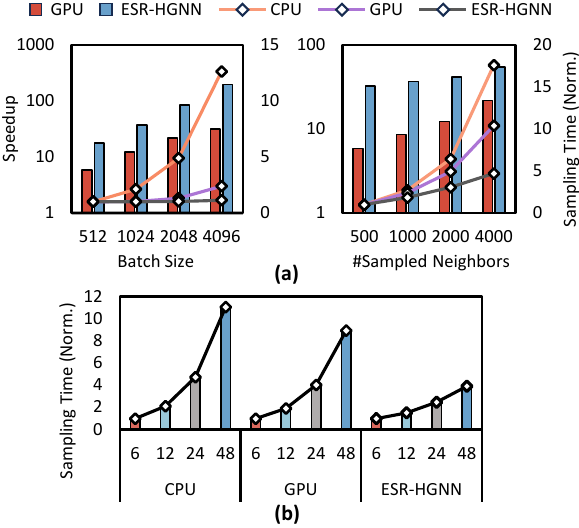}
	\vspace{-4pt}
\caption{Exploration on MAG dataset: (a) Sensitivity regarding batch size and number of sampled neighbors for each target vertex; (b) Scalability across various number of metapaths.}
        \vspace{-6pt}
	\label{fig:exploration}
\end{figure}

We conduct exploratory experiments on two key parameters in metapath-based sampling: batch size and the number of sampled neighbors on the largest dataset MAG, as shown in Fig.\ref{fig:exploration}(a), where the bar chart represents the speedup, and the line chart represents the proportion of sampling time. Overall, ESR-HGNN is more sensitive to the number of sampled neighbors, with the sampling time increasing more significantly as the number of sampled neighbors grows. This is because a higher number of sampled neighbors directly results in a multiplicative increase in traversal paths. It also increases the size of each reusable result block and thus the pressure on the capacity-bounded \textit{Semantic Paths Cache}. When the cache overflows, FIFO replacement writes older result blocks back to HBM; later reuse then incurs a sequential HBM refill rather than repeated random traversal of the matched prefix. Accordingly, as the sampled-neighbor count increases from 500 to 4000, the speedup of ESR-HGNN over the GPU decreases from 4.09$\times$ to 2.43$\times$, while its speedup over the CPU continues to increase. The narrowed advantage over the GPU reflects both the additional result refills and the improved utilization of GPU threads at larger fanouts. In contrast, ESR-HGNN exhibits limited sensitivity to variations in batch size. This can be attributed to its adoption of a vertex-wise sampling paradigm as the underlying processing occurs on a per-vertex basis. For each channel, the \textit{Semantic Paths Cache} only needs to retain the completed, reusable metapath results of the current target vertex; increasing the batch size does not increase this peak resident working set when the fanout is fixed. The sampling time varies very little as long as the total number of target vertices remains the same.

In Fig.~\ref{fig:exploration}(b), we increase the number of metapaths to investigate the scalability of ESR-HGNN on larger datasets. The experimental results reveal that, compared with both the CPU and GPU, the sampling time of ESR-HGNN grows at a considerably slower rate as the number of metapaths increases. Consequently, a higher number of metapaths in larger datasets results in a more pronounced speedup. This is attributed to the introduction of additional semantic redundancy, which facilitates greater reuse of traversal paths. These findings underscore the excellent scalability of ESR-HGNN.

\section{{Discussion}}

\textbf{Near-Memory Processing.} Processing-in-Memory (PIM) and Near-Memory Processing (NMP) reduce the data-movement cost of each remaining adjacency access~\cite{Tesseract,MetaNMP}, whereas ESR-HGNN reduces the number of irregular accesses by eliminating redundant prefix traversals. Redundancy-aware sampling reduces DRAM accesses by 71.56\% on average (Fig.~\ref{fig:incremental_study}(b)). The two approaches are therefore complementary, although their integration requires coordinating trie and cache states across memory partitions. 

\textbf{Higher-Bandwidth HBM.} We use HBM1.0 to align with HiHGNN~\cite{HiHGNN} and the validated HBM1.0 timing configuration in our Ramulator-based evaluation. A sensitivity study using Ramulator~2.1~\cite{ramulator-2-1} shows that HBM2 and HBM3 reduce the normalized sampling time of ESR-HGNN to 0.77$\times$ and 0.69$\times$, respectively. The nonlinear, diminishing gains arise from hop-wise dependence, limited request-level parallelism, random-access latency, bank contention, and limited request coalescing. Nevertheless, newer HBM generations do not affect ESR-HGNN's access-reduction mechanism, as it eliminates redundant prefix traversals and thus reduces the number of DRAM requests regardless of the service rate of the remaining requests.

\section{Conclusion}

This work accelerates mini-batch HGNN inference by eliminating semantic redundancy, proposing a redundancy-aware sampling paradigm to reuse traversal paths and designing a multi-channel sampling unit named ESR-HGNN. A reusability-driven semantic grouping method is also introduced to effectively grouping the metapaths. Experiments show significant improvements in both sampling and end-to-end mini-batch inference performance.

\bibliographystyle{IEEEtranS}
\bibliography{refs}

@inproceedings{HyGCN,
  title={Hygcn: A gcn accelerator with hybrid architecture},
  author={Yan, Mingyu and Deng, Lei and Hu, Xing and others},
  booktitle={2020 IEEE International Symposium on High Performance Computer Architecture (HPCA)},
  pages={15--29},
  year={2020},
  organization={IEEE}
}

@inproceedings{SeHGNN,
  title={Simple and Efficient Heterogeneous Graph Neural Network},
  author={Yang, Xiaocheng and Yan, Mingyu and Pan, Shirui and Ye, Xiaochun and Fan, Dongrui},
  booktitle={Proceedings of the AAAI Conference on Artificial Intelligence},
  volume={37},
  year={2023}
}

@inproceedings{MAGNN,
  title={Magnn: Metapath aggregated graph neural network for heterogeneous graph embedding},
  author={Fu, Xinyu and Zhang, Jiani and Meng, Ziqiao and King, Irwin},
  booktitle={Proceedings of The Web Conference 2020},
  pages={2331--2341},
  year={2020}
}

@inproceedings{HAN,
  title={Heterogeneous graph attention network},
  author={Wang, Xiao and Ji, Houye and Shi, Chuan and others},
  booktitle={The world wide web conference},
  pages={2022--2032},
  year={2019}
}

@ARTICLE{HPN,
  author={Ji, Houye and Wang, Xiao and Shi, Chuan and Wang, Bai and Yu, Philip S.},
  journal={IEEE Transactions on Knowledge and Data Engineering}, 
  title={Heterogeneous Graph Propagation Network}, 
  year={2023},
  volume={35},
  number={1},
  pages={521-532}
}

@inproceedings{Simple-HGN,
  title={Are we really making much progress? Revisiting, benchmarking and refining heterogeneous graph neural networks},
  author={Lv, Qingsong and Ding, Ming and Liu, Qiang and others},
  booktitle={Proceedings of the 27th ACM SIGKDD Conference on Knowledge Discovery \& Data Mining},
  pages={1150--1160},
  year={2021}
}

@article{understand_HGNN,
  title={Characterizing and Understanding HGNNs on GPUs},
  author={Yan, Mingyu and Zou, Mo and Yang, Xiaocheng and others},
  journal={IEEE Computer Architecture Letters},
  volume={21},
  number={2},
  pages={69--72},
  year={2022},
  publisher={IEEE}
}

@inproceedings{7pj,
  title={Highlights of the high-bandwidth memory (hbm) standard},
  author={O’Connor, Mike},
  booktitle={Memory forum workshop},
  volume={3},
  year={2014}
}

@article{ramulator,
  title={Ramulator: A fast and extensible DRAM simulator},
  author={Kim, Yoongu and Yang, Weikun and Mutlu, Onur},
  journal={IEEE Computer architecture letters},
  volume={15},
  number={1},
  pages={45--49},
  year={2015},
  publisher={IEEE}
}

@misc{ramulator-2-1,
      title={Ramulator 2.1: A Composable Memory System Simulator for Modern DRAM Systems}, 
      author={Haocong Luo and F. Nisa Bostancı and Ataberk Olgun and Maria Makeenkova and Ziad Malik and Ipek Akdeniz and Onur Mutlu},
      year={2026},
      eprint={2606.13844},
      archivePrefix={arXiv},
      primaryClass={cs.AR},
      url={https://arxiv.org/abs/2606.13844}, 
}

@inproceedings{igcn,
author = {Geng, Tong and Wu, Chunshu and Zhang, Yongan and others},
title = {I-GCN: A Graph Convolutional Network Accelerator with Runtime Locality Enhancement through Islandization},
year = {2021},
isbn = {9781450385572},
publisher = {Association for Computing Machinery},
address = {New York, NY, USA},
booktitle = {MICRO-54: 54th Annual IEEE/ACM International Symposium on Microarchitecture},
pages = {1051–1063},
numpages = {13},
location = {Virtual Event, Greece},
series = {MICRO '21}
}

@inproceedings{DGL,
  title={Deep graph library: Towards efficient and scalable deep learning on graphs},
  author={Wang, Minjie Yu},
  booktitle={ICLR workshop on representation learning on graphs and manifolds},
  year={2019}
}

@inproceedings{Tesseract,
  title={A scalable processing-in-memory accelerator for parallel graph processing},
  author={Ahn, Junwhan and Hong, Sungpack and Yoo, Sungjoo and Mutlu, Onur and Choi, Kiyoung},
  booktitle={Proceedings of the 42nd Annual International Symposium on Computer Architecture},
  pages={105--117},
  year={2015}
}

@INPROCEEDINGS{technology_scale, 
author={O. {Villa} and D. R. {Johnson} and M. {Oconnor} and others}, 
booktitle={SC '14: Proceedings of the International Conference for High Performance Computing, Networking, Storage and Analysis}, 
title={Scaling the Power Wall: A Path to Exascale}, 
year={2014}, 
volume={}, 
number={}, 
pages={830-841}, 
doi={10.1109/SC.2014.73}, 
ISSN={2167-4337}, 
month={Nov},}

@online{CACTI,
	title = {CACTI},
	url = {http://www.hpl.hp.com/research/cacti/},
	urldate = {2018-09-11}
}

@INPROCEEDINGS{ReGNN,
  author={Chen, Cen and Li, Kenli and Li, Yangfan and Zou, Xiaofeng},
  booktitle={2022 IEEE International Symposium on High-Performance Computer Architecture (HPCA)}, 
  title={ReGNN: A Redundancy-Eliminated Graph Neural Networks Accelerator}, 
  year={2022},
  volume={},
  number={},
  pages={429-443},
  doi={10.1109/HPCA53966.2022.00039}}

@INPROCEEDINGS {FlowGNN,
author = {R. Sarkar and S. Abi-Karam and Y. He and L. Sathidevi and C. Hao},
booktitle = {2023 IEEE International Symposium on High-Performance Computer Architecture (HPCA)},
title = {FlowGNN: A Dataflow Architecture for Real-Time Workload-Agnostic Graph Neural Network Inference},
year = {2023},
volume = {},
issn = {},
pages = {1099-1112},
doi = {10.1109/HPCA56546.2023.10071015},
publisher = {IEEE Computer Society},
address = {Los Alamitos, CA, USA},
month = {mar}
}

@ARTICLE{GRIP,
  author={Kiningham, Kevin and Levis, Philip and Ré, Christopher},
  journal={IEEE Transactions on Computers}, 
  title={GRIP: A Graph Neural Network Accelerator Architecture}, 
  year={2023},
  volume={72},
  number={4},
  pages={914-925},
  doi={10.1109/TC.2022.3197083}}

@inproceedings{GNNSampler,
  title={GNNSampler: Bridging the gap between sampling algorithms of GNN and hardware},
  author={Liu, Xin and Yan, Mingyu and Song, Shuhan and Lv, Zhengyang and Li, Wenming and Sun, Guangyu and Ye, Xiaochun and Fan, Dongrui},
  booktitle={Joint European Conference on Machine Learning and Knowledge Discovery in Databases},
  pages={498--514},
  year={2022},
  organization={Springer}
}

@inproceedings{GROW,
  author       = {Ranggi Hwang and
                  Minhoo Kang and
                  Jiwon Lee and
                  Dongyun Kam and
                  Youngjoo Lee and
                  Minsoo Rhu},
  title        = {{GROW:} {A} Row-Stationary Sparse-Dense {GEMM} Accelerator for Memory-Efficient
                  Graph Convolutional Neural Networks},
  booktitle    = {{IEEE} International Symposium on High-Performance Computer Architecture,
                  {HPCA} 2023, Montreal, QC, Canada, February 25 - March 1, 2023},
  pages        = {42--55},
  publisher    = {{IEEE}},
  year         = {2023},
  bibsource    = {dblp computer science bibliography, https://dblp.org}
}

@inproceedings{REFLIP-HUAKE,
  author       = {Yu Huang and
                  Long Zheng and
                  Pengcheng Yao and
                  others},
  title        = {Accelerating Graph Convolutional Networks Using Crossbar-based Processing-In-Memory
                  Architectures},
  booktitle    = {{IEEE} International Symposium on High-Performance Computer Architecture,
                  {HPCA} 2022, Seoul, South Korea, April 2-6, 2022},
  pages        = {1029--1042},
  publisher    = {{IEEE}},
  year         = {2022},
  bibsource    = {dblp computer science bibliography, https://dblp.org}
}

@inproceedings{MetaNMP, 
author = {Chen, Dan and He, Haiheng and Jin, Hai and others}, title = {MetaNMP: Leveraging Cartesian-Like Product to Accelerate HGNNs with Near-Memory Processing}, year = {2023}, isbn = {9798400700958}, publisher = {Association for Computing Machinery}, address = {New York, NY, USA}, booktitle = {Proceedings of the 50th Annual International Symposium on Computer Architecture}, articleno = {56}, numpages = {13}, location = {Orlando, FL, USA}, series = {ISCA '23} }

@article{GraphSage,
  title={Inductive representation learning on large graphs},
  author={Hamilton, Will and Ying, Zhitao and Leskovec, Jure},
  journal={Advances in neural information processing systems},
  volume={30},
  year={2017}
}

@ARTICLE{Comprehensive_Survey_GNN_Distributed_Training,
  author={Lin, Haiyang and Yan, Mingyu and Ye, Xiaochun and Fan, Dongrui and Pan, Shirui and Chen, Wenguang and Xie, Yuan},
  journal={Proceedings of the IEEE}, 
  title={A Comprehensive Survey on Distributed Training of Graph Neural Networks}, 
  year={2023},
  volume={111},
  number={12},
  pages={1572-1606}}

@article{MultiGCN,
  title={Multi-Node Acceleration for Large-Scale GCNs},
  author={Sun, Gongjian and Yan, Mingyu and Wang, Duo and others},
  journal={IEEE Transactions on Computers},
  volume={71},
  number={12},
  pages={3140--3152},
  year={2022},
  publisher={IEEE}
}

@ARTICLE{HiHGNN,
  author={Xue, Runzhen and Han, Dengke and Yan, Mingyu and Zou, Mo and Yang, Xiaocheng and Wang, Duo and Li, Wenming and Tang, Zhimin and Kim, John and Ye, Xiaochun and Fan, Dongrui},
  journal={IEEE Transactions on Parallel and Distributed Systems}, 
  title={HiHGNN: Accelerating HGNNs Through Parallelism and Data Reusability Exploitation}, 
  year={2024},
  volume={35},
  number={7},
  pages={1122-1138}
}

@article{understand_hgnn_training,
author = {Han, Dengke and Yan, Mingyu and Ye, Xiaochun and Fan, Dongrui},
title = {Characterizing and Understanding HGNN Training on GPUs},
year = {2025},
issue_date = {March 2025},
publisher = {Association for Computing Machinery},
address = {New York, NY, USA},
volume = {22},
number = {1},
issn = {1544-3566},
url = {https://doi.org/10.1145/3703356},
doi = {10.1145/3703356},
journal = {ACM Trans. Archit. Code Optim.},
month = mar,
articleno = {9},
numpages = {25}
}

@inproceedings{HetGNN,
author = {Zhang, Chuxu and Song, Dongjin and Huang, Chao and Swami, Ananthram and Chawla, Nitesh V.},
title = {Heterogeneous Graph Neural Network},
year = {2019},
isbn = {9781450362016},
publisher = {Association for Computing Machinery},
address = {New York, NY, USA},
booktitle = {Proceedings of the 25th ACM SIGKDD International Conference on Knowledge Discovery \& Data Mining},
pages = {793–803},
numpages = {11},
location = {Anchorage, AK, USA},
series = {KDD '19}
}

@article{weibo-recommendation,
  title={Context Aware Sentiment Link Prediction in Heterogeneous Social Network},
  author={Anping Zhao and Yu Yu},
  journal={Cognitive Computation},
  year={2021},
  volume={14},
  pages={300 - 309},
}

@INPROCEEDINGS{dataset-recommendation,
  author={Altaf, Basmah and Akujuobi, Uchenna and Yu, Lu and Zhang, Xiangliang},
  booktitle={2019 IEEE International Conference on Data Mining (ICDM)}, 
  title={Dataset Recommendation via Variational Graph Autoencoder}, 
  year={2019},
  volume={},
  number={},
  pages={11-20}}

@inproceedings{eda_1,
author = {Zhou, Xinyi and Ye, Junjie and Pui, Chak-Wa and Shao, Kun and Zhang, Guangliang and Wang, Bin and Hao, Jianye and Chen, Guangyong and Heng, Pheng Ann},
title = {Heterogeneous Graph Neural Network-Based Imitation Learning for Gate Sizing Acceleration},
year = {2022},
isbn = {9781450392174},
publisher = {Association for Computing Machinery},
address = {New York, NY, USA},
booktitle = {Proceedings of the 41st IEEE/ACM International Conference on Computer-Aided Design},
articleno = {57},
numpages = {9},
location = {San Diego, California},
series = {ICCAD '22}
}

@article{eda_2,
  title={Versatile multi-stage graph neural network for circuit representation},
  author={Yang, Shuwen and Yang, Zhihao and Li, Dong and Zhang, Yingxueff and Zhang, Zhanguang and Song, Guojie and Hao, Jianye},
  journal={Advances in Neural Information Processing Systems},
  volume={35},
  pages={20313--20324},
  year={2022}
}

@inproceedings{tencent_malware_detection,
author = {Ye, Yanfang and Hou, Shifu and Chen, Lingwei and Lei, Jingwei and Wan, Wenqiang and Wang, Jiabin and Xiong, Qi and Shao, Fudong},
title = {Out-of-sample node representation learning for heterogeneous graph in real-time android malware detection},
year = {2019},
isbn = {9780999241141},
publisher = {AAAI Press},
booktitle = {Proceedings of the 28th International Joint Conference on Artificial Intelligence},
pages = {4150–4156},
numpages = {7},
location = {Macao, China},
series = {IJCAI'19}
}

@inproceedings{abnormal_event_detection,
author = {Fan, Shaohua and Shi, Chuan and Wang, Xiao},
title = {Abnormal Event Detection via Heterogeneous Information Network Embedding},
year = {2018},
isbn = {9781450360142},
publisher = {Association for Computing Machinery},
address = {New York, NY, USA},
booktitle = {Proceedings of the 27th ACM International Conference on Information and Knowledge Management},
pages = {1483–1486},
numpages = {4},
location = {Torino, Italy},
series = {CIKM '18}
}

@ARTICLE{sampling_survey,
  author={Liu, Xin and Yan, Mingyu and Deng, Lei and Li, Guoqi and Ye, Xiaochun and Fan, Dongrui},
  journal={IEEE/CAA Journal of Automatica Sinica}, 
  title={Sampling Methods for Efficient Training of Graph Convolutional Networks: A Survey}, 
  year={2022},
  volume={9},
  number={2},
  pages={205-234}}

@inproceedings{metapath2vec,
author = {Dong, Yuxiao and Chawla, Nitesh V. and Swami, Ananthram},
title = {metapath2vec: Scalable Representation Learning for Heterogeneous Networks},
year = {2017},
isbn = {9781450348874},
publisher = {Association for Computing Machinery},
address = {New York, NY, USA},
booktitle = {Proceedings of the 23rd ACM SIGKDD International Conference on Knowledge Discovery and Data Mining},
pages = {135–144},
numpages = {10},
location = {Halifax, NS, Canada},
series = {KDD '17}
}

@article{randomwalk,
  title={Random walks on graphs},
  author={Lov{\'a}sz, L{\'a}szl{\'o}},
  journal={Combinatorics, Paul erdos is eighty},
  volume={2},
  number={1-46},
  pages={4},
  year={1993}
}

@article{GNN_Characterization_Survey,
  title={Survey on Characterizing and Understanding GNNs from a Computer Architecture Perspective},
  author={Wu, Meng and Yan, Mingyu and Li, Wenming and Ye, Xiaochun and Fan, Dongrui and Xie, Yuan},
  journal={IEEE Transactions on Parallel and Distributed Systems},
  year={2025},
  publisher={IEEE}
}

@article{GraphAgile,
  title={Graphagile: An fpga-based overlay accelerator for low-latency gnn inference},
  author={Zhang, Bingyi and Zeng, Hanqing and Prasanna, Viktor K},
  journal={IEEE Transactions on Parallel and Distributed Systems},
  volume={34},
  number={9},
  pages={2580--2597},
  year={2023},
  publisher={IEEE}
}

@ARTICLE{ViTeGNN,
  author={Zhou, Hongkuan and Zhang, Bingyi and Kannan, Rajgopal and Busart, Carl and Prasanna, Viktor K.},
  journal={IEEE Transactions on Parallel and Distributed Systems}, 
  title={ViTeGNN: Towards Versatile Inference of Temporal Graph Neural Networks on FPGA}, 
  year={2025},
  volume={36},
  number={3},
  pages={502-519},
  doi={10.1109/TPDS.2024.3521897}}

@article{louvain,
  title={Fast unfolding of communities in large networks},
  author={Blondel, Vincent D and Guillaume, Jean-Loup and Lambiotte, Renaud and Lefebvre, Etienne},
  journal={Journal of statistical mechanics: theory and experiment},
  volume={2008},
  number={10},
  pages={P10008},
  year={2008},
  publisher={IOP Publishing}
}

@inproceedings{openhgnn,
author = {Han, Hui and Zhao, Tianyu and Yang, Cheng and Zhang, Hongyi and Liu, Yaoqi and Wang, Xiao and Shi, Chuan},
title = {OpenHGNN: An Open Source Toolkit for Heterogeneous Graph Neural Network},
year = {2022},
isbn = {9781450392365},
publisher = {Association for Computing Machinery},
address = {New York, NY, USA},
booktitle = {Proceedings of the 31st ACM International Conference on Information \& Knowledge Management},
pages = {3993–3997},
numpages = {5},
location = {Atlanta, GA, USA},
series = {CIKM '22}
}

@misc{DropOut,
      title={Improving neural networks by preventing co-adaptation of feature detectors}, 
      author={Geoffrey E. Hinton and Nitish Srivastava and Alex Krizhevsky and Ilya Sutskever and Ruslan R. Salakhutdinov},
      year={2012},
      eprint={1207.0580},
      archivePrefix={arXiv},
      primaryClass={cs.NE}
}

@inproceedings{DropNode,
author = {Liu, Meng and Gao, Hongyang and Ji, Shuiwang},
title = {Towards Deeper Graph Neural Networks},
year = {2020},
isbn = {9781450379984},
publisher = {Association for Computing Machinery},
address = {New York, NY, USA},
url = {https://doi.org/10.1145/3394486.3403076},
doi = {10.1145/3394486.3403076},
booktitle = {Proceedings of the 26th ACM SIGKDD International Conference on Knowledge Discovery \& Data Mining},
pages = {338–348},
numpages = {11},
location = {Virtual Event, CA, USA},
series = {KDD '20}
}

@inproceedings{ADE-HGNN,
author = {Han, Dengke and Wu, Meng and Xue, Runzhen and Yan, Mingyu and Ye, Xiaochun and Fan, Dongrui},
title = {ADE-HGNN: Accelerating HGNNs Through Attention Disparity Exploitation},
year = {2024},
isbn = {978-3-031-69765-4},
publisher = {Springer-Verlag},
address = {Berlin, Heidelberg},
booktitle = {Euro-Par 2024: Parallel Processing: 30th European Conference on Parallel and Distributed Processing, Madrid, Spain, August 26–30, 2024, Proceedings, Part II},
pages = {91–106},
numpages = {16},
location = {Madrid, Spain}
}

@inproceedings{GDR-HGNN,
author = {Xue, Runzhen and Yan, Mingyu and Han, Dengke and Teng, Yihan and Tang, Zhimin and Ye, Xiaochun and Fan, Dongrui},
title = {GDR-HGNN: A Heterogeneous Graph Neural Networks Accelerator Frontend with Graph Decoupling and Recoupling},
year = {2024},
isbn = {9798400706011},
publisher = {Association for Computing Machinery},
address = {New York, NY, USA},
booktitle = {Proceedings of the 61st ACM/IEEE Design Automation Conference},
articleno = {4},
numpages = {6},
location = {San Francisco, CA, USA},
series = {DAC '24}
}

@INPROCEEDINGS{TLV-HGNN,
  author={Han, Dengke and Wang, Duo and Yan, Mingyu and Ye, Xiaochun and Fan, Dongrui},
  booktitle={2025 IEEE 43rd International Conference on Computer Design (ICCD)}, 
  title={TLV-HGNN: Thinking Like a Vertex for Memory-Efficient HGNN Inference}, 
  year={2025},
  volume={},
  number={},
  pages={730-737},
  doi={10.1109/ICCD65941.2025.00110}}

@ARTICLE{CoGNN,
  author={Zhong, Kai and Zeng, Shulin and Hou, Wentao and Dai, Guohao and Zhu, Zhenhua and Zhang, Xuecang and Xiao, Shihai and Yang, Huazhong and Wang, Yu},
  journal={IEEE Transactions on Computer-Aided Design of Integrated Circuits and Systems}, 
  title={CoGNN: An Algorithm-Hardware Co-Design Approach to Accelerate GNN Inference With Minibatch Sampling}, 
  year={2023},
  volume={42},
  number={12},
  pages={4883-4896},}

\section{Biography Section}

\vspace{-30pt}
\begin{IEEEbiography}[{\includegraphics[width=1in,height=1.25in,clip,keepaspectratio]{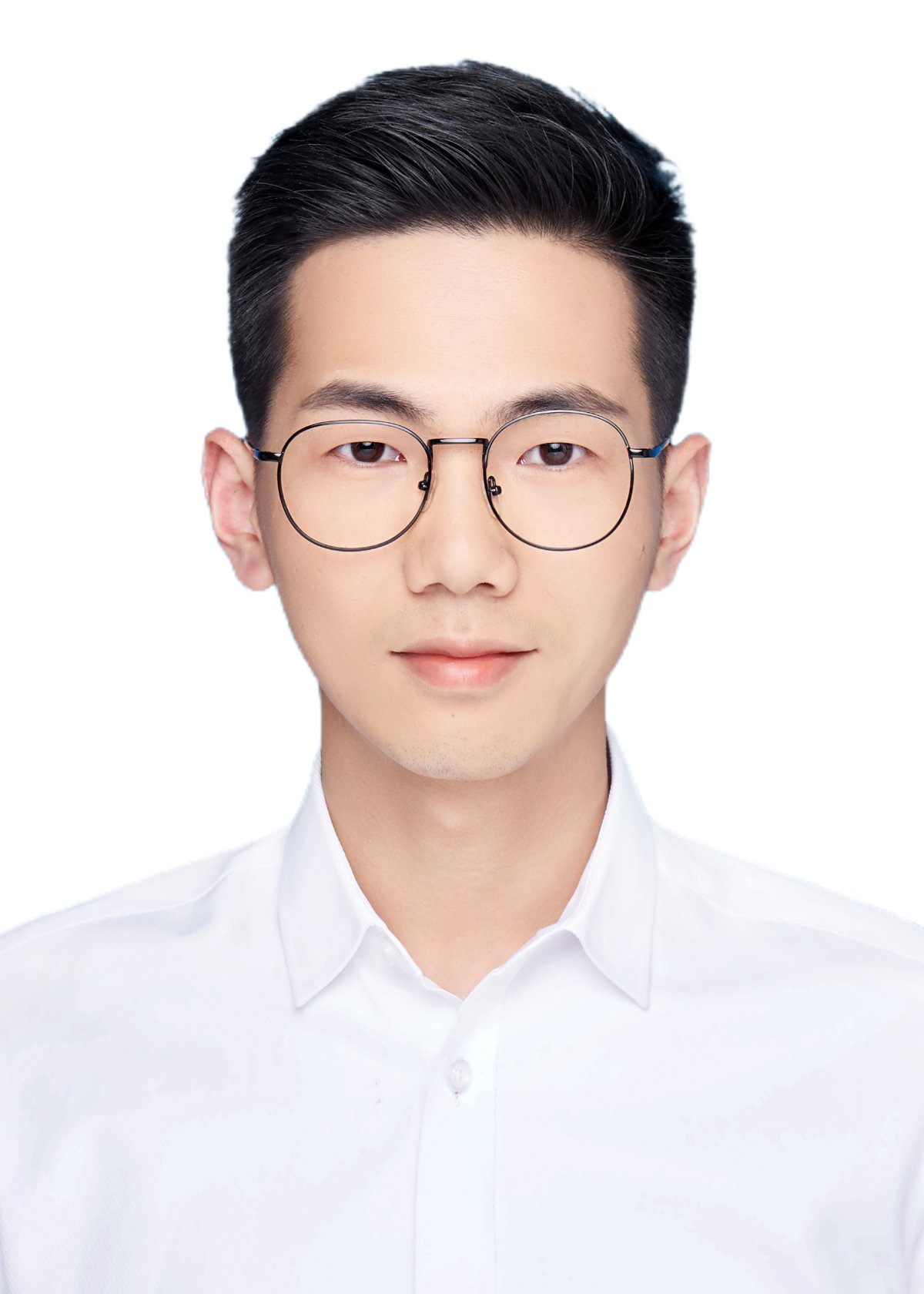}}]{Dengke Han} 
He is currently a Ph.D. candidate at the Institute of Computing Technology, Chinese Academy of Sciences, Beijing, China. His research interests include graph-based hardware accelerators, algorithm performance analysis and optimization, and high-throughput computer architecture.
\end{IEEEbiography}

\vspace{-30pt}
\begin{IEEEbiography}[{\includegraphics[width=1in, height=1.25in, clip, keepaspectratio]{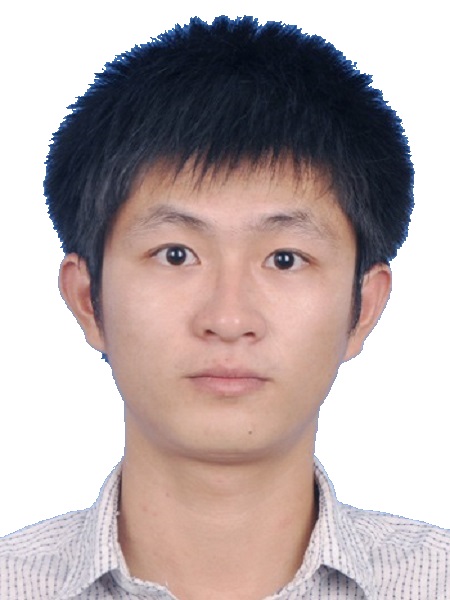}}] {Mingyu Yan} 
received his Ph.D. degree from University of Chinese Academy of Sciences, Beijing, China in 2020. He is currently an associate professor in Institute of Computing Technology, Chinese Academy of Sciences, Beijing, China. His current research interests include graph processing algorithm, graph-based hardware accelerator, and high-throughput computer architecture.
\end{IEEEbiography}

\vspace{-30pt}
\begin{IEEEbiography}[{\includegraphics[width=1in,height=1.25in,clip,keepaspectratio]{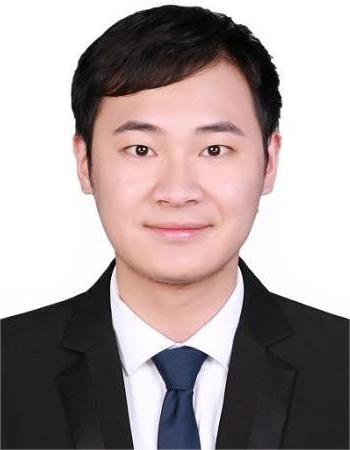}}]{Duo Wang}
received his Ph.D. degree from University of Chinese Academy of Sciences, Beijing, China in 2024. His current research interests include processor design space exploration, high-performance computer architecture and software simulation.
\end{IEEEbiography}

\vspace{-30pt}
\begin{IEEEbiography}[{\includegraphics[width=1in, height=1.25in, clip, keepaspectratio]{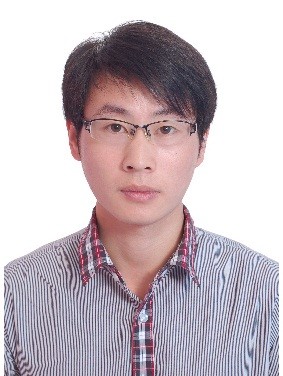}}] {Wenming Li} 
received the Ph.D. degree in computer architecture from Institute of Computing Technology, Chinese Academy of Sciences, Beijing, in 2016. He is currently an associate professor in Institute of Computing Technology, Chinese Academy of Sciences, Beijing. His main research interests include high-throughput processor architecture, dataflow architecture and software simulation. 
\end{IEEEbiography}

\vspace{-35pt}
\begin{IEEEbiography}[{\includegraphics[width=1in, height=1.25in, clip, keepaspectratio]{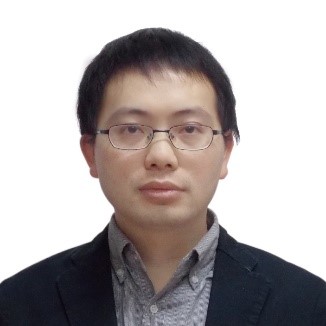}}] {Xiaochun Ye} 
received his Ph.D. degree in computer architecture from Institute of Computing Technology, Chinese Academy of Sciences, Beijing, in 2010. He is currently a professor and Ph.D. supervisor in Institute of Computing Technology, Chinese Academy of Sciences, Beijing. His main research interests include high-performance computer architecture and software simulation.
\end{IEEEbiography}

\vspace{-40pt}
\begin{IEEEbiography}[{\includegraphics[width=1in, height=1.25in, clip, keepaspectratio]{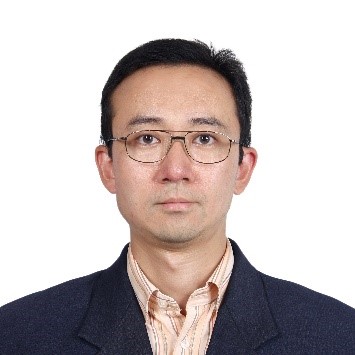}}] {Dongrui Fan} 
received his Ph.D. degree in computer architecture from Institute of Computing Technology, Chinese Academy of Sciences, Beijing, in 2005. He is currently a professor and Ph.D. supervisor in Institute of Computing Technology, Chinese Academy of Sciences, Beijing. His main research interests include high-throughput computer architecture and high-performance computer architecture. 
\end{IEEEbiography}

\end{document}